\documentclass[twocolumn]{aastex631}

\usepackage{graphicx}	
\usepackage{amsmath}	
\usepackage{natbib}
\usepackage{multirow} 
\usepackage{booktabs} 
\usepackage{float} 
\usepackage{placeins} 
\usepackage{enumerate}
\usepackage{fancyvrb} 

\newcommand{\angstrom}{\textup{\AA}} 

\def\heasarc{{HEASARC}}

\def\xmm{\textit{XMM-Newton}}
\def\pn{{EPIC-pn}}
\def\om{{OM}}

\def\rosat{\textit{ROSAT}}
\def\pspc{{PSPC}}
\def\swift{\textit{Swift}}
\def\iue{\textit{IUE}}
\def\feka{{Fe~K$\alpha$}}

\def\xspec{\texttt{XSPEC}}
\def\pexrav{\texttt{pexrav}}

\def\po{\texttt{powerlaw}}

\def\bb{\texttt{bbody}}

\def\cpl{\texttt{\texttt{cutoffpl}}}
\def\zphabs{\texttt{zphabs}}
\def\phabs{\texttt{phabs}}
\def\zxipcf{\texttt{zxipcf}}
\def\relxilllp{\texttt{relxilllp}}

\def\comptt{\texttt{compTT}}
\def\asurv{\texttt{ASURV}}
\def\topcat{\texttt{TOPCAT}}

\begin{document}

\title{A UV to X-ray view of soft excess in type 1 AGNs: I. sample selection and spectral profile}
\shorttitle{Soft X-ray Excess, Hot Corona, and Accretion Disk}
\shortauthors{Chen et al.}

\correspondingauthor{Shi-Jiang Chen \& Jun-Xian Wang}\email{JohnnyCsj666@gmail.com, jxw@ustc.edu.cn}

\author[0009-0004-9950-9807]{Shi-Jiang Chen}
\affiliation{School of Astronomy and Space Science, University of Science and Technology of China, Hefei 230026, People's Republic of China}

\author[0000-0002-4419-6434]{Jun-Xian Wang}
\affiliation{School of Astronomy and Space Science, University of Science and Technology of China, Hefei 230026, People's Republic of China}
\affiliation{CAS Key Laboratory for Research in Galaxies and Cosmology, Department of Astronomy, University of Science and Technology of China, Hefei, Anhui 230026, People's Republic of China}

\author[0000-0003-2280-2904]{Jia-Lai Kang}
\affiliation{School of Astronomy and Space Science, University of Science and Technology of China, Hefei 230026, People's Republic of China}
\affiliation{CAS Key Laboratory for Research in Galaxies and Cosmology, Department of Astronomy, University of Science and Technology of China, Hefei, Anhui 230026, People's Republic of China}

\author[0000-0003-2573-8100]{Wen-Yong Kang}
\affiliation{School of Astronomy and Space Science, University of Science and Technology of China, Hefei 230026, People's Republic of China}
\affiliation{CAS Key Laboratory for Research in Galaxies and Cosmology, Department of Astronomy, University of Science and Technology of China, Hefei, Anhui 230026, People's Republic of China}

\author[0000-0002-9265-2772]{Hao Sou}
\affiliation{School of Astronomy and Space Science, University of Science and Technology of China, Hefei 230026, People's Republic of China}
\affiliation{CAS Key Laboratory for Research in Galaxies and Cosmology, Department of Astronomy, University of Science and Technology of China, Hefei, Anhui 230026, People's Republic of China}

\author[0000-0002-2941-6734]{Teng Liu}
\affiliation{School of Astronomy and Space Science, University of Science and Technology of China, Hefei 230026, People's Republic of China}
\affiliation{CAS Key Laboratory for Research in Galaxies and Cosmology, Department of Astronomy, University of Science and Technology of China, Hefei, Anhui 230026, People's Republic of China}

\author[0000-0002-4223-2198]{Zhen-Yi Cai}
\affiliation{School of Astronomy and Space Science, University of Science and Technology of China, Hefei 230026, People's Republic of China}
\affiliation{CAS Key Laboratory for Research in Galaxies and Cosmology, Department of Astronomy, University of Science and Technology of China, Hefei, Anhui 230026, People's Republic of China}

\author[0000-0001-8515-7338]{Zhen-Bo Su}
\affiliation{School of Astronomy and Space Science, University of Science and Technology of China, Hefei 230026, People's Republic of China}
\affiliation{CAS Key Laboratory for Research in Galaxies and Cosmology, Department of Astronomy, University of Science and Technology of China, Hefei, Anhui 230026, People's Republic of China}

\begin{abstract}
A core sample of 59 unobscured type 1 AGNs with simultaneous \xmm~X-ray and UV observations is compiled from archive to probe the nature of soft X-ray excess (SE). In the first paper of this series, our focus centers on scrutinizing the spectral profile of the soft excess. Of the sources, $\approx 71\%~(42/59)$ exhibit powerlaw-like (po-like) soft excess, while $\approx 29\%~(17/59)$ exhibit blackbody-like (bb-like) soft excess. We show a cut-off powerlaw could uniformly characterize both types of soft excesses, with median $E_\text{cut}$ of 1.40 keV for po-like and 0.14 keV for bb-like. For the first time, we report a robust and quantitative correlation between the SE profile and SE strength (the ratio of SE luminosity to that of the primary powerlaw continuum in 0.5 -- 2.0 keV), indicating that stronger soft excess is more likely to be po-like, or effectively has a higher $E_\text{cut}$. This correlation cannot be explained by ionized disk reflection alone, which produces mostly bb-like soft excess ($E_\text{cut}\sim 0.1\ \text{keV}$) as revealed by \relxilllp~simulation. Remarkably, we show with simulations that a toy hybrid scenario, where both ionized disk reflection (\relxilllp, with all reflection parameters fixed at default values except for ionization of the disk) and warm corona (\comptt, with temperature fixed at 1 keV) contribute to the observed soft excess, can successfully reproduce the observed correlation. This highlights the ubiquitous hybrid nature of the soft X-ray excess in AGNs, and underscores the importance of considering both components while fitting the spectra of soft excess. 
\end{abstract}

\keywords{Galaxies: active -- Galaxies: nuclei -- X-rays: galaxies}

\section{Introduction}\label{intro}
Being among the most powerful sources in the universe, Active Galactic Nuclei (AGN) consistently captivate astronomers due to their spectral complexity and the uncertainties surrounding their nature \citep[e.g.,][]{Cai&Wang2023}. In X-ray observations, a powerlaw-shaped primary continuum, typically accompanied by iron emission lines and a reflection continuum, predominates the hard X-ray band ($> 2\  \text{keV}$). This primary continuum is widely believed to originate from inverse-Compton processes, where a fraction of disk photons are up-scattered through interactions with a central hot corona \citep[e.g.,][]{Haardt&Maraschi1991, Haardt&Maraschi1993}. The iron lines (and the reflection continuum) are produced when a fraction of Comptonized photons illuminate different regions of the accretion disk \citep[e.g.,][]{Buchner+2014, Bambi+2021}, either the outer parts (narrow \feka~line, e.g., \citealt{Patrick+2012}), or the close proximity to the central black hole (broad \feka~line, e.g., \citealt{Reynolds2003, Falocco+2012}). 

However, the soft X-ray band tells a more complex story. When extrapolating the primary continuum above 2 keV to the softer band,  a discernible excess is observed in over half of the radio-quiet type 1 AGN population \citep{Bianchi+2009}. This ``soft X-ray excess" has long been recognized to exhibit diverse spectral profiles \citep[e.g.,][]{Gierlinski&Done2004,Piconcelli+2005,Grupe+2010}, resembling either a blackbody (hereafter bb-like) or a powerlaw (hereafter po-like).

The physical origin of the soft excess remains an open question, with two prevailing theories -- the ``warm corona" and ``ionized disk reflection" -- currently leading the efforts to elucidate its nature. In the first scenario, a warm, optically-thick plasma, termed ``warm corona", is believed to boost a certain fraction of disk photons to the soft X-ray band, thereby creating the observed soft excess \citep{Magdziarz+1998,Mehdipour+2011,Petrucci+2013,Rozanska2015,Petrucci+2018,Kawanaka&Mineshige2024}. Another variation of this scenario suggests that the gravitational energy in the innermost part of the accretion disk is divided between powering the warm corona (producing the soft X-ray excess) and the hot corona (producing the primary continuum) \citep[e.g.,][]{Done+2012,Jin+2012a,Kubota&Done2018}. In the second scenario, a relativistically blurred reflection feature adds an additional component in the soft X-ray band when the inner region of the ionized disk is irradiated by a lamp-post hot corona \citep[e.g.,][]{Ross&Fabian1993,Ballantyne+2001,Miniutti&Fabian2004,Ross&Fabian2005,Crummy+2005,Merloni+2006,Garcia&Kallman2010,Dovciak+2011,Bambi+2021}.

A general approach to understanding the nature of the soft excess involves fitting the spectra with physically-motivated models \citep[e.g.,][]{Dewangan+2007,Fabian+2012,Chiang+2015,Jin+2016,Jiang+2018,Tripathi+2019,Middei+2020,Xu+2021a,Xu+2021,Chalise+2022,Waddell+2023}. While these models often yield good fitting statistics, they face persistent challenges. Typically, fitting the spectra of sources with a strong soft excess using only ionized reflection models requires extreme physical parameter configurations, including a high spin value ($a_*\gtrsim0.993$), a very low coronal height ($h\sim R_\mathrm{Hor}$), and a very high disk density ($n_e\gtrsim10^{18}\ \mathrm{cm}^{-3}$) \citep[e.g.][]{Boissay+2014,Garcia+2019,Xu+2021}. Sometimes fitting with reflection models results in residuals in the hard X-ray band \citep[e.g.,][]{Liu+2020}. On the other hand, in the case of a pure warm corona, coronal ionization equilibrium (CIE) predicts prominent absorption lines due to photoelectric absorption \citep{Garcia+2019}, which contrasts with the smooth soft excess observed in most AGNs. Recent findings suggest that a scenario where both warm corona and ionized reflection coexist has the potential to alleviate these observational challenges \citep[e.g.,][]{Porquet+2018,Ballantyne2020,Petrucci+2020,Ballantyne&Xiang2020,Porquet+2021,Xiang+2022,Ballantyne+2024}. 

Observationally, it has long been standard practice to characterize the soft excess in a large sample using a single family of phenomenological model, i.e. either blackbody or powerlaw \citep[e.g.,][]{Crummy+2005, Boissay+2016,Gliozzi&Williams2020,Waddell&Gallo2020,Ding+2022, Piconcelli+2005,Liu+2022, Waddell+2023}. However, it has been pointed out that neither a blackbody nor a powerlaw is able to fully  parameterize the soft excess emission in all PG quasars \citep{Piconcelli+2005}, indicating the absence of a universal spectral shape for the soft excess. Interestingly, the theoretical work of \cite{Xiang+2022} predicted that when both ionized reflection and a warm corona contribute,  the resulting soft excess will exhibit a wide variety of spectral profiles. 
Therefore, it is essential to extensively examine the variation of the spectral profile of the soft excess across a large sample and in individual sources. This approach could uncover and probe the co-existence of the two components (i.e., the warm corona and the ionized reflection), providing deeper insights into the physical mechanisms driving the soft excess.

Furthermore, since in the ``warm corona" model the soft excess is tightly connected to UV emission \citep[e.g.,][]{Petrucci+2013}, incorporating UV data alongside X-ray observations is necessary for these studies. For instance, previous research, utilizing \rosat~\pspc~(0.1-2.4 keV) and \iue~data, has reported a strong link between the soft excess and the accretion disk \citep{Walter&Fink1993,Liu&Qiao2010}. Nevertheless, comprehensive investigations that consider both spectral profiles and broadband correlations are still scarce.

In this series of papers, we aim to probe the nature of the soft excess using the high-quality X-ray spectra of XMM \pn~and simultaneous UV data recorded by XMM \om~for a large sample of type 1 AGNs, addressing the following questions:
\begin{enumerate}
    \item How does the spectral shape of soft excess vary within a large sample, and how does it depend on broadband SED and other parameters? 

    \item How does the observed soft excess strength correlate with broadband emission and other physical parameters in a large sample? 

    \item How does the soft excess vary in individual sources? How are their variations coordinated with those of broadband emission? 
    
\end{enumerate}

This work (Paper I) is structured as follows. In \S \ref{SampDR}, we introduce our sample and provide a detailed overview of the \xmm~data reduction process. The methods for spectral fitting and the criteria for selecting the core sample are outlined in \S \ref{SpecMod}. To provide a direct view of the soft excess profile, we present the data-to-model ratio, as well as the ``unfolded" soft excess spectrum in \S \ref{Profile}. In \S \ref{Simulation}, we offer a quantitative assessment of the relationship between soft excess shape and strength in the core sample, alongside comparisons with ionized disk reflection and the double-component scenario. We also compare the soft excess shape with other broadband parameters (UV-to-Xray luminosity ratio, Eddington ratio, and primary continuum photon index) and give discussion.

Hereafter, we shall use ``primary continuum" and ``PC" interchangeably, both referring to the primary X-ray powerlaw continuum. The terms ``soft X-ray excess", ``soft excess", and ``SE" will be used synonymously to indicate the excess component in the soft X-ray band (0.5 -- 2 keV)\footnote{Note that, as we will show below, in a few sources, the soft excess component may extend beyond 2 keV. However, the contribution above 2 keV to the total soft excess flux remains small, even in the most extreme cases ($<$ 30\%). Therefore, to maintain consistency with the literature, we continue to adopt the 0.5 –- 2 keV range for measuring the soft excess flux. }. To convert flux into luminosity, we assume a cosmology with $H_0 = 70\ \text{km}\ \text{s}^{-1}\ \text{Mpc}^{-1},\ \Omega_\text{m} = 0.3$ and $\Omega_\Lambda = 0.7$.

\section{Sample and data reduction} \label{SampDR}
\subsection{The initial sample} \label{sample}
Bright and unobscured X-ray spectrum is crucial for successfully separating the soft excess from the primary powerlaw continuum. To construct a large sample of X-ray bright type 1 AGNs (Seyfert 1 and QSO 1), we consider two well-known catalogs: the \swift-BAT 105-month catalog (BAT105, \citealt{Koss+2017,Oh+2018}) and the Catalog of AGN in the \xmm~Archive (CAIXA, \citealt{Bianchi+2009}), which include a total of 452 sources. To ensure that the soft X-ray excess is effectively covered by  the \pn\ bandpass (0.5 -- 10 keV), and to minimize the bias introduced in K-correction to UV data, we restrict our analysis to a subset of 428 sources in the local universe ($z<0.4$). Additionally, we exclude 66 sources located near the Galactic equator ($N_\text{H,Gal}>10^{21}\ \text{cm}^{-2}$)\footnote{The $N_\text{H,Gal}$ values are obtained using the tool ``\texttt{NH}'' provided by NASA’s High Energy Astrophysics Science Archive Research Center (\heasarc).} to avoid strong Galactic absorption to UV and soft X-ray.

The \om~comprises three UV filters (UVW2, UVM2, UVW1) with effective wavelengths at $2120\text{\AA}$, $2310\text{\AA}$, and $2910\text{\AA}$, respectively, along with three optical filters (U, V, B) at $3440\text{\AA}$, $4500\text{\AA}$, and $5430\text{\AA}$, respectively \citep[e.g.,][]{Page+2012, Liu+2021}. Compared to the optical filters, the UV filters are better suited to our objectives because they measure the emission from the relatively inner accretion disk, and are less prone to contamination from host galaxies \citep[e.g.,][]{Grupe+2010}. Among the three UV filters, UVW1 has a peak effective area approximately twice as large as UVM2 and around eight times as large as UVW2 \citep{Page+2012}. Moreover, UVW1 is more frequently utilized in XMM observations. For instance, in the \om~Serendipitous Ultraviolet source survey catalogue (SUSS 6.0, \citealt{Page+2012}), approximately 49\% of sources have UVW1 exposures, compared to only around 10\% for UVM2 and 5\% for UVW2. Therefore in this work we only consider XMM exposures with simultaneous \pn\ and UVW1 observations. After cross-matching with the \xmm~Master Log \& Public Archive\footnote{https://heasarc.gsfc.nasa.gov/W3Browse/xmm-newton/xmmmaster.html}, we compile a catalog of 151 sources (totaling 471 observations) with effective UVW1 photometry.

\subsection{\om~data reduction} \label{om}
We reduce the \om~observation data file (ODF) of each observation with the pipeline \texttt{omichain} in \xmm~Science Analysis System (SAS, version 20.0.0), to retrieve a combined list of sources detected in the UVW1 band during each exposure. From the source list, we pick the one closest to the coordinates of our target (provided by NASA/IPAC Extragalactic Database) within 5\arcsec. To convert UVW1 count rate ($\text{ct}\ \text{s}^{-1}$) to monochromatic flux ($\text{erg}\ \text{cm}^{-2}\ \text{s}^{-1}\ \text{\AA}^{-1}$), we apply a conversion factor of $4.76 \times 10^{-16}$ (see the \xmm~SAS user guide).

For our sample, we account for Galactic dust extinction and perform a K-correction. The Galactic extinction is expressed as $A(\text{UVW1}) = R(\text{UVW1}) \times E\text{(B-V)}$, where the $E\text{(B-V)}$ for each source is obtained from \citealt{Schlegel+1998}, and the UVW1 extinction coefficient $R(\text{UVW1})$ is taken as 5.28 \citep{Page+2012}. To perform K-correction, we adopt a universal UV spectral slope $\alpha=0.65$, assuming the UV SED can be described as $F_{\nu} \sim \nu^{-\alpha}$ \citep{Natali+1998}. Consequently, the corrected monochromatic UVW1 flux at the rest frame in our sample is calculated as: 
\begin{equation}
    \centering
    F_\text{UVW1,int} = F_\text{UVW1,obs} \times 10^{0.4 \times R(\text{UVW1}) \times E\text{(B-V)}} \times (1+z)^{\alpha-1}
\end{equation}

Finally, to facilitate direct comparison with studies in literature, we convert the observed monochromatic UVW1 flux ($2910\angstrom$) to monochromatic $2500\angstrom$ flux, assuming the same spectral slope $\alpha=0.65$. The X-ray monochromatic flux to be used in the following sections is obtained from the unfolded spectra, based on model 3 defined in \S \ref{model}. 

We note that while host galaxies can contribute over $\sim 50\%$ of the total light in optical or longer wavelength bands, their contamination in the UV for nearby Seyfert I galaxies is significantly less pronounced \citep[e.g.,][]{Grupe+2010}. Studies based on UV grism spectroscopy \citep[e.g.,][]{Mehdipour+2015} and high-resolution HST imaging \citep[e.g.,][]{MunozMarin+2007} show that host galaxy contributions in the UV are typically below $\sim 10\%$ for nearby Seyfert I galaxies. Moreover, the standard OM pipeline employs an aperture radius of 2–3 arcseconds, and for $\sim 15\%$ sources in our sample where host galaxy structure is resolved in \om~images, potential contamination from star-forming regions has been further reduced. Therefore, we conclude that host galaxy contamination does not significantly impact our results, and no further correction is applied.

\subsection{\pn~data reduction} \label{pn}
The \pn~data are processed using \xmm~SAS, employing the current calibration files (CCF 3.13). High background intervals are filtered out, and the source extraction regions are optimized utilizing the SAS task \texttt{eregionanalyse}. The background spectra are extracted from source-free regions. A detailed processing procedure is illustrated in \citealt{Kang&Wang2024}.

The SAS task \texttt{epatplot} is employed to assess the potential pile-up effect. Pile-up is considered severe and non-negligible if the observed-to-model fraction of single events ($s$) is significantly smaller than one\footnote{https://heasarc.gsfc.nasa.gov/docs/xmm/sas/help/epatplot.pdf}. Among our sample, 73 observations exhibit pronounced pile-up ($1-s > 3\sigma$, where $\sigma$ is the error of $s$). Consequently, for these observations, we utilize an annular region to extract the source spectrum. We maintain the outer radius fixed at the value determined by \texttt{eregionanalyse}, while iteratively testing inner radii of $5''$, $15''$, and $25''$, until the pile-up effect is mitigated ($1-s<3\sigma$).

For a robust constraint on the soft X-ray excess parameters, a reliable estimation of the hard X-ray primary continuum is crucial. Therefore, we retain only observations with more than 50 spectral bins in the 2.5 -- 10.0 keV range, after rebinning the \pn~spectrum to ensure a minimum of 25 photons per bin. This step reduces the sample to 127 sources with a total of 451 observations.

\section{X-ray Spectral Fitting} \label{SpecMod}
\subsection{Spectral models} \label{model}
\begin{figure*}
    \centering
    \includegraphics[width=2.0\columnwidth]{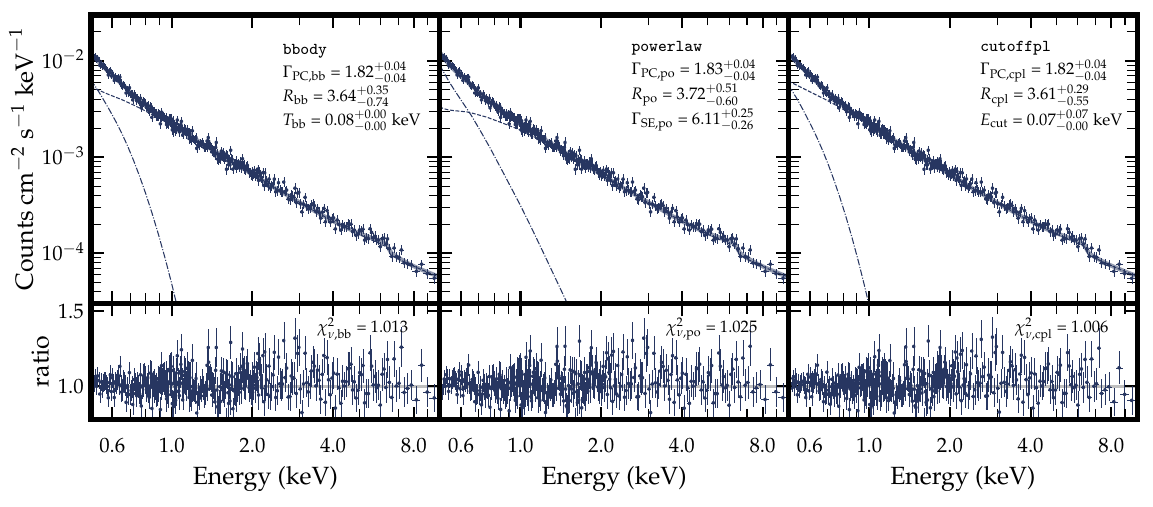}
    \caption{An example of bb-like soft excess (1H 0419-577, OBSID:0148000601). 
    The data, divided by the response effective area at each energy channel, is shown along with the best-fit folded models (left to right: model 1, 2, and 3). The profile of bb-like soft excess is generally constrained below 1 keV. When fitted with cut-off powerlaw (middle), $E_\text{cut}\sim 0.1\ \text{keV}$.}
    \label{fig:eg_bblike}
\end{figure*}
\begin{figure*}
    \centering
    \includegraphics[width=2.0\columnwidth]{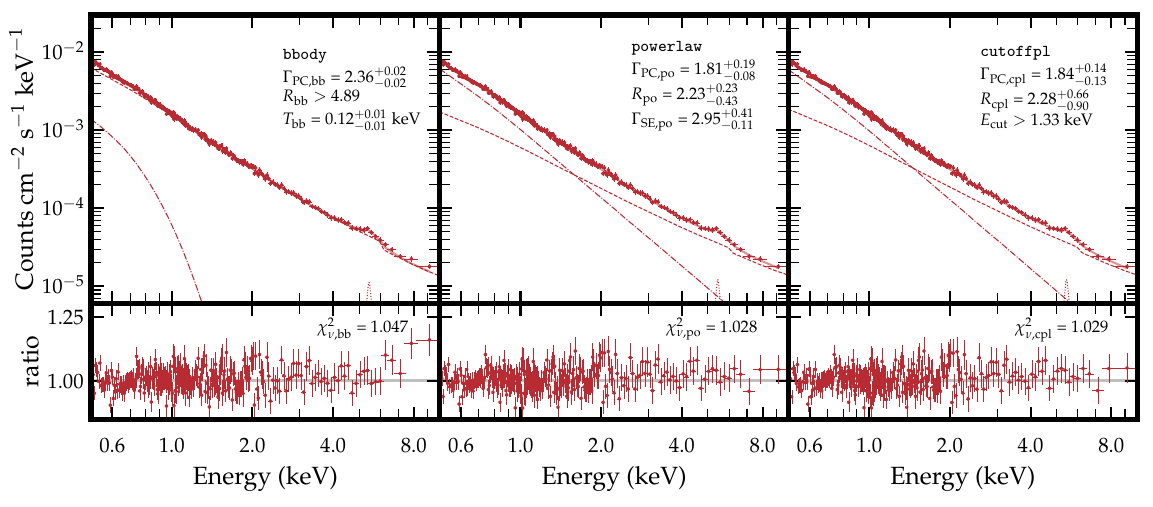}
    \caption{Similar to Fig. \ref{fig:eg_bblike}, but for an example of po-like soft excess (PG 1116+215, OBSID:0201940101). The profile of po-like soft excess is generally more extended, and has non-negligible contribution above 1 keV. Note an excess between data and folded model at $\sim 7\ \text{keV}$ when the soft excess is modeled with blackbody. The cut-off powerlaw behaves like a normal powerlaw with similar fitting statistics ($\chi^2=1009.3$ for cut-off powerlaw, and $1009.4$ for powerlaw).} 
    \label{fig:eg_polike}
\end{figure*}

We perform X-ray spectra fitting using \xspec~\citep{Arnaud1996}, within 0.5 -- 10 keV band (termed the broad X-ray band). Three phenomenological models are applied to the broad X-ray band spectra, which in \xspec~terminology are: 
\begin{verbatim}
1: phabs*zphabs*(zbbody+pexrav+zgauss1+zgauss2) 
    
2: phabs*zphabs*(zpowerlw+pexrav+zgauss1+zgauss2)

3: phabs*zphabs*(zcutoffpl+pexrav+zgauss1+zgauss2)
\end{verbatim}

The best-fit reduced chi-squares are denoted as $\chi^2_{\nu,\text{bb}}$ (model 1), $\chi^2_{\nu,\text{po}}$ (model 2), and $\chi^2_{\nu,\text{cpl}}$ (model 3), respectively. Common to all three models, \pexrav~\citep{Magdziarz&Zdziarski1995} represents the contribution from both primary continuum and distant neutral reflection. We fixed the parameters of \pexrav~at their default values except for the photon index $\Gamma$, flux (normalization), and reflection fraction $R$ (with a hard upper limit of 5). We note that the effective \xmm~band has limited power in constraining $R$, and the inter-observation changes in $R$ for a single source are typically within $2\sigma$ range. Therefore, for sources with multiple observations, we fixed $R$ at the value determined by a link fit. 

To account for the narrow and broad \feka~lines, which are prominent features for AGNs \citep[e.g.,][]{Fabian+1989, Patrick+2012, Falocco+2012}, two Gaussian components (\texttt{zgauss1} and \texttt{zgauss2}) are included in the models. For the narrow core, the rest-frame centroid energy \texttt{LineE} and line width \texttt{Sigma} are fixed at 6.4 keV and 0.019 keV, respectively \citep[e.g.,][]{Shu+2010, Kang2020}. For the broad line, we set a hard limit of 5 -- 6.5 keV for \texttt{LineE} and 0 -- 1 keV for \texttt{Sigma}. Two layers of photoelectric absorber are applied to model the intrinsic and Galactic neutral absorption. The $N_\text{H}$ value is fixed for the Galactic absorption (\phabs) while set free for the intrinsic one (\zphabs).

We aim to conduct a uniform analysis of the statistical properties of the soft X-ray excess without presuming its physical origin \textit{a priori}, i.e. whether it arises from a warm corona or the disk reflection process. To this end, we employ three phenomenological models to characterize the soft X-ray excess: blackbody (\bb, bb), powerlaw (\po, po), and cut-off powerlaw (\cpl, cpl). While the first two models have been extensively utilized in prior studies of the soft X-ray excess \citep[e.g.,][]{Bianchi+2009, Gliozzi&Williams2020, Waddell&Gallo2020, Ding+2022, Liu+2022} due to their simplicity and effective characterization, we also explore the possibility of a cut-off powerlaw for the following reasons:
\begin{enumerate}
    \item Firstly, the shape of soft X-ray excess demonstrates great diversity, encompassing both bb-like (defined as $\chi^2_{\text{bb}}-\chi^2_{\text{po}}<0$) and po-like ($\chi^2_{\text{bb}}-\chi^2_{\text{po}}>0$) profiles \citep[e.g.][]{Gierlinski&Done2004,Piconcelli+2005,Grupe+2010,Waddell+2023}. The distinct characteristics at ``high-energy" end ($1\sim 2$ keV) between blackbody and powerlaw (blackbody drops much faster than powerlaw, at an exponential rate), may suggest different underlying physics for the two types. Refer to Appendix \ref{profile-abs} for a detailed discussion on the potential effects of X-ray absorption on the identification of soft excess profile. Due to the presence of both types of soft X-ray excesses, relying solely on either blackbody or powerlaw may be inadequate to describe the shape and origin of the soft excess for the entire population. 
    
    \item Secondly, a cut-off powerlaw, which retains the powerlaw characteristics at the soft tail, while bends down exponentially at the hard tail, has the potential to describe both types of soft X-ray excess (also applied in \citealt{Matt+2014,Petrucci+2020} to fit soft excess). The cut-off energy, $E_\text{cut}$, serves as the key parameter controlling the profile. Specifically, when $E_\text{cut}\sim 0.1\ \text{keV}$, the cut-off powerlaw resembles a blackbody in the \pn~band\footnote{Note that in this case, the lower energy powerlaw component of a cut-off powerlaw spectrum falls below the effective range of \pn. Consequently, its photon index is poorly constrained or carries no physical meanings. For further discussion, see \S \ref{TabKeyPar}.}; while $E_\text{cut}\gtrsim1\ \text{keV}$ results in a profile similar to a normal powerlaw (see Appendix \ref{cosample}). Therefore, the flexibility of the cut-off powerlaw allows for effective characterization of both bb-like and po-like sources in the sample, as well as the relation between soft X-ray excess profile (reflected by $E_\text{cut}$) and other physical quantities.
    
    \item For our sample, the cut-off powerlaw yields good fit. To preliminarily exclude observations affected by warm or partial absorption, we focus on a subset of 246 observations (hereafter the preliminary sample), where at least one of three models yields a relatively good fit ($\chi^2_{\nu,\text{bb}}<1.2$ or $\chi^2_{\nu,\text{cpl}}<1.2$ or $\chi^2_{\nu,\text{po}}<1.2$). The median value for reduced chi-square ($\chi^2_{\nu}$) is $1.058$ for powerlaw, $1.055$ for blackbody, while $1.038$ for cut-off powerlaw. Regarding the improvement of chi-square ($\Delta\chi^2$), the overall improvement achieved by model 3 is modest relative to model 1 and 2 across the entire sample, with a median $\chi^2_\text{bb}-\chi^2_\text{cpl}$ of $5.98$ for $\chi^2_\text{bb}-\chi^2_\text{cpl}$, corresponding to an F-test significance\footnote{calculated utilizing the median degree of freedom and chi-square of the sample, hereafter the same.} of $0.02$, and a median $\chi^2_\text{po}-\chi^2_\text{cpl}$ of 2.95, with an F-test significance of $0.08$. However, the cut-off powerlaw (model 3) significantly outperforms the blackbody model (model 1) for the po-like sources, with a median $\chi^2_\text{bb}-\chi^2_\text{cpl}$ of 13.71 (F-test significance $3\times 10^{-4}$), while achieving a comparable fit to bb-like sources (median $\chi^2_\text{bb}-\chi^2_\text{cpl}$ = 0.17). Similarly, the cut-off powerlaw model achieves a substantial improvement over the powerlaw model for bb-like sources, with a median $\chi^2_\text{po}-\chi^2_\text{cpl}$ is $37.06$ (F-test significance $2\times 10^{-9}$), while achieving a comparable fit for po-like sources (median $\chi^2_\text{pl}-\chi^2_\text{cpl}$ = $1.20$). Therefore, the cut-off powerlaw model serves as a consistently superior option across the entire sample, delivering a more robust fit than either the blackbody or powerlaw models alone.
\end{enumerate}

To illustrate the two types of soft excess profiles and demonstrate the capability of the cut-off powerlaw in describing both, we present in Fig. \ref{fig:eg_bblike} and Fig. \ref{fig:eg_polike} the spectrum data (divided by the response effective area) of a typical po-like source and a bb-like source, along with best fit models and corresponding data-to-model ratios. The bb-like soft excess shows a rapid decline above $\sim 1\ \text{keV}$, whereas the po-like soft excess is more extended and exhibits contributions above $\sim 1\ \text{keV}$. Additionally, in the left panel of Fig. \ref{fig:eg_polike}, it is evident that fitting a broad po-like soft excess with a blackbody model would erroneously result in a steeper primary continuum and large residuals above $\sim 7\ \text{keV}$. Overall, for both bb-like and po-like soft excess profiles, the cut-off powerlaw model can achieve fits similar to or even better than those obtained with the blackbody or powerlaw models.

\subsection{Selection of the core sample} \label{cosample}
To conduct a robust statistical analysis on the properties of soft excess, it is essential to use a core sample devoid of significant warm or neutral absorption. Warm absorption poses a substantial challenge in accurately determining the profile and luminosity of the soft excess (see Appendix \ref{profile-abs} for detailed discussion), and fitting complex absorption can introduce significant model-dependent uncertainties. We note the criterion used in the selection of the preliminary sample is rather rudimentary; therefore, a more systematic and detailed selection process is necessary, as outlined below.

Starting from the preliminary sample of 246 observations:
\begin{enumerate}
    \item We first drop spectra with best-fit \zphabs~column density (from model 3) larger than $10^{21}\ \text{cm}^{-2}$. This step removes 37 observations.\\
    
    \item For the remaining spectra, we employ a test model:
    \begin{verbatim}
 phabs*zphabs*zxipcf*(zcutoffpl+pexrav
                     +zgauss1+zgauss2)
    \end{verbatim}
    in which the \texttt{zxipcf} component models a partially covered ionized absorber. We exclude observations where adding the \texttt{zxipcf} component improves the spectral fitting by $\Delta\chi^2>10$. This step eliminates 86 observations, resulting in a sample of 123 observations (from 59 sources).
\end{enumerate}

\begin{figure*}
    \centering
    \includegraphics[width=2.0\columnwidth]{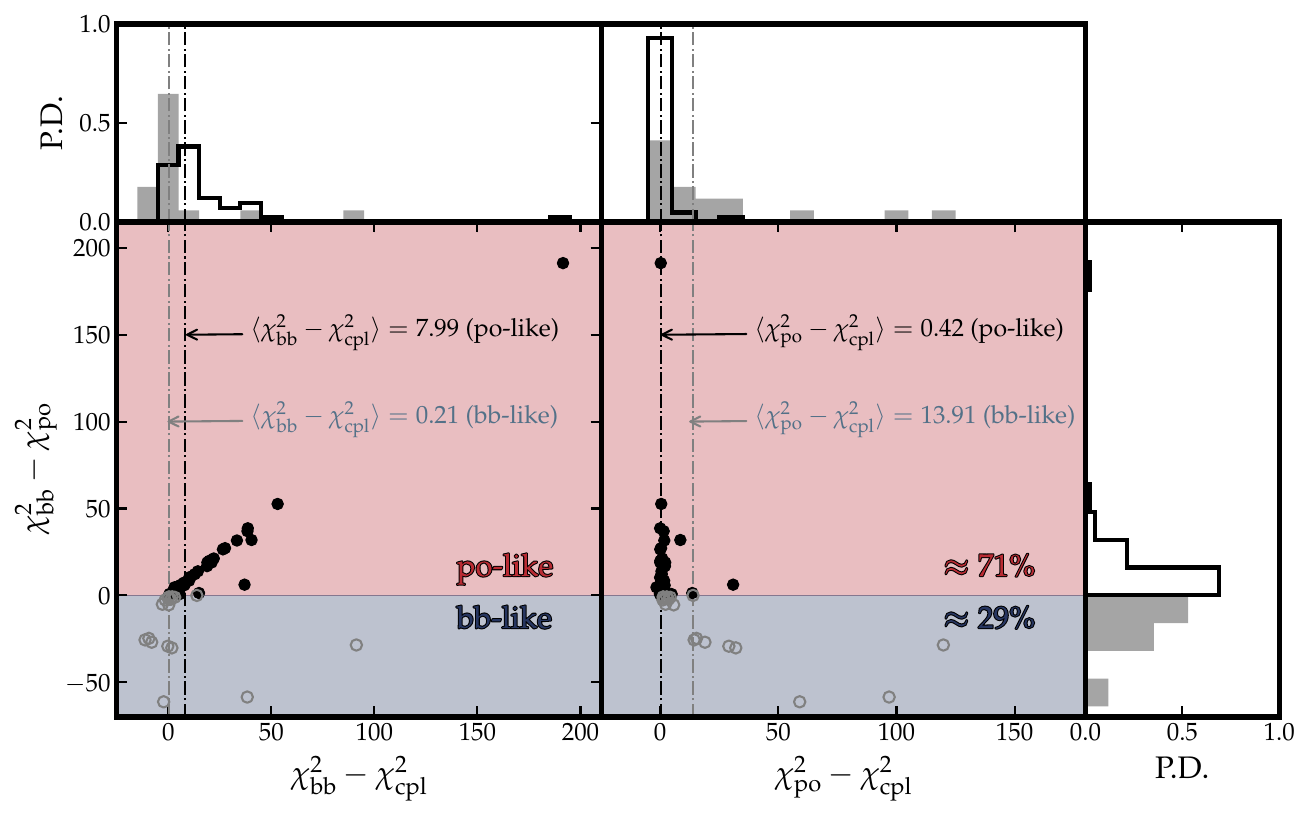}
    \caption{The comparison of 
    chi-square ($\chi^2$) when fitting with models 1, 2, and 3. 
    The Y-axis of the lower panels shows the difference in $\chi^2$ between model 1 and 2, i.e.,  $\chi^2_{\text{bb}}$-$\chi^2_{\text{po}}$. Po-like sources ($\chi^2_{\text{bb}}$-$\chi^2_{\text{po}}$ $>$ 0) are represented by filled circles, while bb-like sources ($\chi^2_{\text{bb}}$-$\chi^2_{\text{po}}$ $<$ 0) are depicted with open circles. In the left and right panels, the X-axes plot the difference between model 1 and 3 ($\chi^2_{\text{bb}} - \chi^2_{\text{cpl}}$), and model 2 and 3 ($\chi^2_{\text{po}} - \chi^2_{\text{cpl}}$), respectively. Histograms for both axes are included, with separate distributions for po–like and bb-like sources. Median $\chi^2$ improvements in model 3 relative to models 1 and 2 are marked by vertical lines for each source type. 
    }
    \label{fig:chi2_cmp}
\end{figure*}

Apart from warm absorption, some emission lines, either collisionally excited or photoionization excited, may also contribute to the soft X-ray regime \citep[e.g.,][]{Reeves+2016,Porquet+2024a}. These emission lines, typically with FWHM in the range of $\sim$ 100 -- 1000 $\mathrm{km}\ \mathrm{s}^{-1}$, could potentially be associated with outflowing clouds at torus scales \citep[e.g.,][]{Buhariwalla+2023,Buhariwalla+2024}. To investigate the presence of such a component and assess its influence on the soft excess shape, we have added an additional \texttt{zgauss} to model 3, allowing its center to vary within the rest-frame range of 0.5 -- 1.5 $\mathrm{keV}$. The line width is fixed at $0\ \mathrm{eV}$, as the intrinsic width ($\sim \mathrm{eV}$, according to the aforementioned literature) is generally far below PN resolution ($\sim 100\ \mathrm{eV}$). Only $\sim 10\%$ observations show a significant improvement in fit ($\Delta \chi^2>10$), and even in these cases, changes in soft excess parameters (cut-off energy, soft excess strength) are marginal, remaining within $1\sigma$ range. Since the presence of such lines would not alter the results presented in this work, we retain the spectral fitting results of model 3 for further analysis.

Since the primary focus of this study is on the inter-source variation of the soft excess, we only retain the observation with the longest X-ray exposure for sources with multiple \xmm~observations. We defer a dedicated study on the variation of the soft excess in individual sources to Paper III. The final core sample comprises 59 observations from 59 type 1 AGNs (see Table \ref{tab:ParTab2} and \ref{tab:ParTab3} in Appendix \ref{TabKeyPar} for the list of targets/observations and key parameters we derived).

In Fig. \ref{fig:chi2_cmp} we compare the fitting statistics of the three models applied to the core sample. Po-like soft excess constitutes the majority of the core sample ($42/59\approx71\%$, depicted by black filled circles), while the contribution from bb-like soft excess is also notable ($17/59\approx29\%$, illustrated by gray open circles). For po-like sources, the cut-off powerlaw model achieves a fit as good as the powerlaw model, and significantly better than blackbody (with a median $\Delta \chi^2$ of 7.99)\footnote{The exact value for the median $\Delta \chi^2$ differs from that presented in \S\ref{model} because the core sample is analyzed here, instead of the preliminary sample discussed in \S\ref{model}.}. For bb-like sources, the cut-off powerlaw model performs comparably to the blackbody model, and much better than powerlaw (with a median $\Delta \chi^2$ of 13.91).

Considering $E_\text{cut}$ when fitting with model 3, a median value of 0.14 keV is found for bb-like sources. On the other hand, we find that $E_\text{cut}$ is unconstrained for $32/42\approx 76\%$ of po-like sources due to limited spectral quality. Therefore we employ the Kaplan-Meier estimator implemented in the Astronomy SURVival analysis (\asurv, \citealt{Feigelson&Nelson1985,Isobe+1986}) package, and derive a median value of 1.40 keV (see upper panel of Fig. \ref{fig:Ecvschi}). The median $E_\mathrm{cut}$ for the entire core sample is $\sim 1.0\ \mathrm{keV}$.

Finally, for the hard X-ray continuum, our model 3 yields a median photon index of $1.85$ with a sample standard deviation of $0.26$, close aligning with the results from nearby AGN surveys. For instance, \citealt{Piconcelli+2005} reports a median photon index of $1.85$ with a standard deviation $0.27$; \citealt{Vasudevan&Fabian2009} finds a median of $1.84$ and a standard deviation $0.32$; and for the $\log N_\mathrm{H}<21\ \mathrm{cm}^{-2}$ subsample of BAT105 \citep{Ricci+2017a}, a median of $1.87$ and a standard deviation $0.25$ are reported. Similarly, the reflection strength in our sample, with a median of $1.34$ and a sample standard deviation of $1.38$, is also comparable to the measurements from the BAT105 subsample (median $1.40$, standard deviation $2.62$).

\section{The spectral profile of the soft excess} \label{Profile}
\subsection{The data-to-model ratio plot} \label{ratio}
\begin{figure*}
    \centering
    \includegraphics[width=2.0\columnwidth]{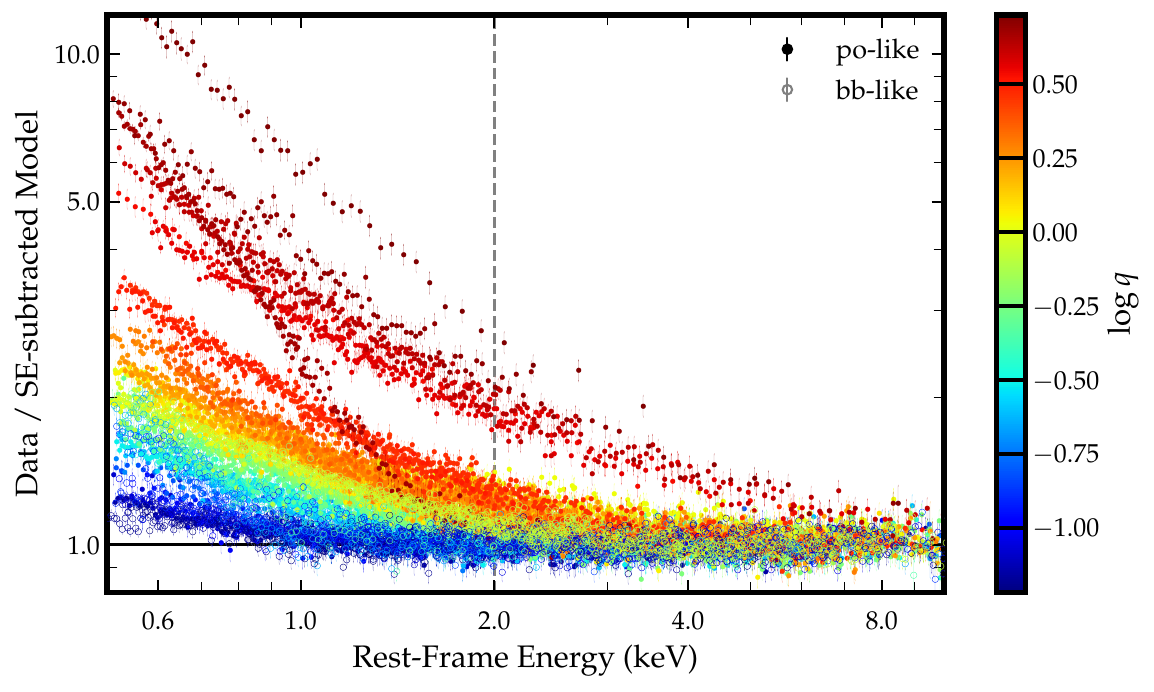}
    \caption{The rest-frame data-to-model ratio plot (using the best-fit model 3, but excluding the soft excess component) of the core sample. Po-like sources are marked with filled circles, while bb-like sources are marked with open circles. Each spectrum is color coded according to the soft X-ray excess strength $\log q$ (see text for definition).}
    \label{fig:DMratio}
\end{figure*}
In Fig. \ref{fig:DMratio} we plot the rest-frame data-to-model ratio (using the best-fit model 3, but excluding the soft excess component) to illustrate the spectral shape of the soft excess of our core sample. This method, compared with plotting the residuals (data minus folded model), highlights the spectral shape of the soft excess without being affected by the energy-dependent instrument response, and shows its strength relative to the underlying hard X-ray powerlaw. Each spectrum is color-coded based on the soft X-ray excess strength in logarithmic scale, $\log q$, defined as the ratio of soft excess luminosity (model 3) and primary continuum luminosity in the 0.5 -- 2 keV band (similar to \citealt{Boissay+2016}, SX1 in \citealt{Gliozzi&Williams2020}):
\begin{equation}
    \log q\equiv\log\frac{L_\text{SE,0.5 -- 2,cpl}}{L_\text{PC,0.5 -- 2,cpl}} \label{q}
\end{equation}
and choosing a different energy range (e.g., 0.5 -- 10 keV) would not change our major conclusion.

We observe a general trend where a relatively stronger soft excess exhibits a more extended shape, with non-negligible contributions above 2 keV, similar to the pattern shown in Fig. 5 of \citealt{Boissay+2016}, which presented the stacked spectra for a couple of AGN groups based on soft excess strengths. In Fig. \ref{fig:DMratio}, we also assign different symbols for po-like sources (filled circles) and bb-like sources (open circles). An interesting observation is that, unlike po-like sources, all bb-like sources are located at the bottom of the figure, indicating weaker soft excess. The evolving trend of the soft excess profile and strength will be further explored in the following sections.

\subsection{The normalized unfolded spectra of the soft excess} \label{ufse}
\begin{figure*}
    \centering    \includegraphics[width=2\columnwidth]{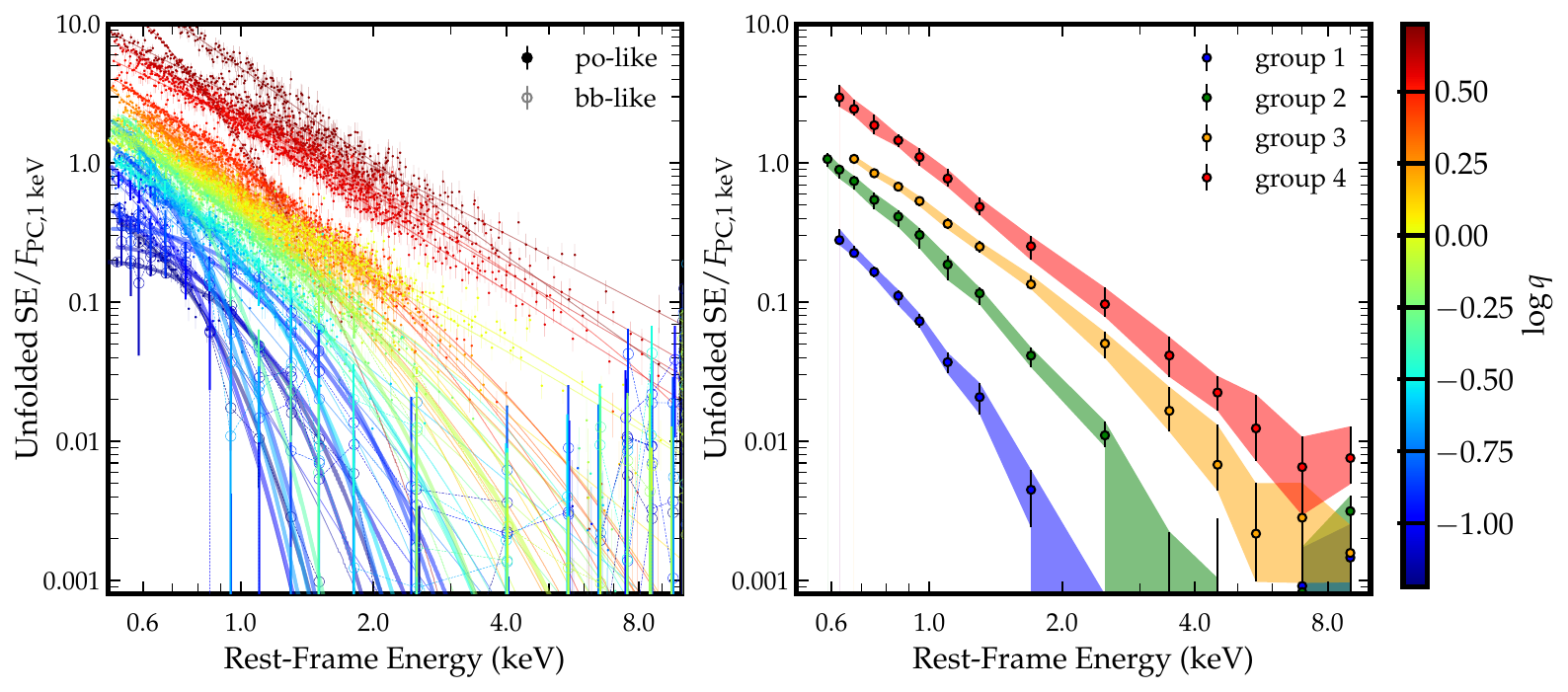}
    \caption{The ``unfolded'' soft excess, normalized by the primary powerlaw continuum flux at 1 keV, is shown for the core sample. The left panel displays the normalized unfolded soft excess for all 59 sources, while the right panel presents a stacked view, with sources grouped into four categories based on soft excess strength. The error bars for the weighted mean stacked spectra are derived using a bootstrapping method applied to the sources within each group. } \label{fig:SEufspec}
\end{figure*}

We note that the spectral shapes presented in Fig. \ref{fig:DMratio} is dependent on the slope of the primary continuum. This means that two spectra with identical soft excess components but different primary continuum slopes could exhibit different soft excess spectral shapes in Fig. \ref{fig:DMratio}. To illustrate the spectral shape of the soft excess component more accurately, we present in Fig. \ref{fig:SEufspec} the ``unfolded" soft excess spectra. These spectra are obtained by subtracting the primary continuum and \feka~model from the unfolded spectra and then normalizing by the rest-frame 1.0 keV flux of the underlying primary continuum ($F_\text{PC, 1 keV}$).

The left panel of Fig. \ref{fig:SEufspec} shows the rest-frame unfolded spectra of the soft excess for all 59 sources in the core sample. These spectra have been corrected for neutral absorption and are color-coded similarly to Fig. \ref{fig:DMratio}. To enhance clarity, we evenly divide the core sample into four groups based on soft excess strength $q$, and display in the right panel of Fig. \ref{fig:SEufspec} the stacked unfolded spectra of soft excess for each group.

A rapid drop-off toward higher energy, indicative of a bb-like profile, is noticeable in the soft excess profiles with weaker soft excess (group 1 and 2). Conversely, as the strength increases, the soft excess profile extends and resembles a powerlaw spectrum. This observation aligns with the conclusion drawn in \S \ref{ratio}.

\section{Quantitative assessment of soft excess profile: from data to simulations}
\label{Simulation}
\begin{figure}
    \centering
    \includegraphics[width=1.0\columnwidth]{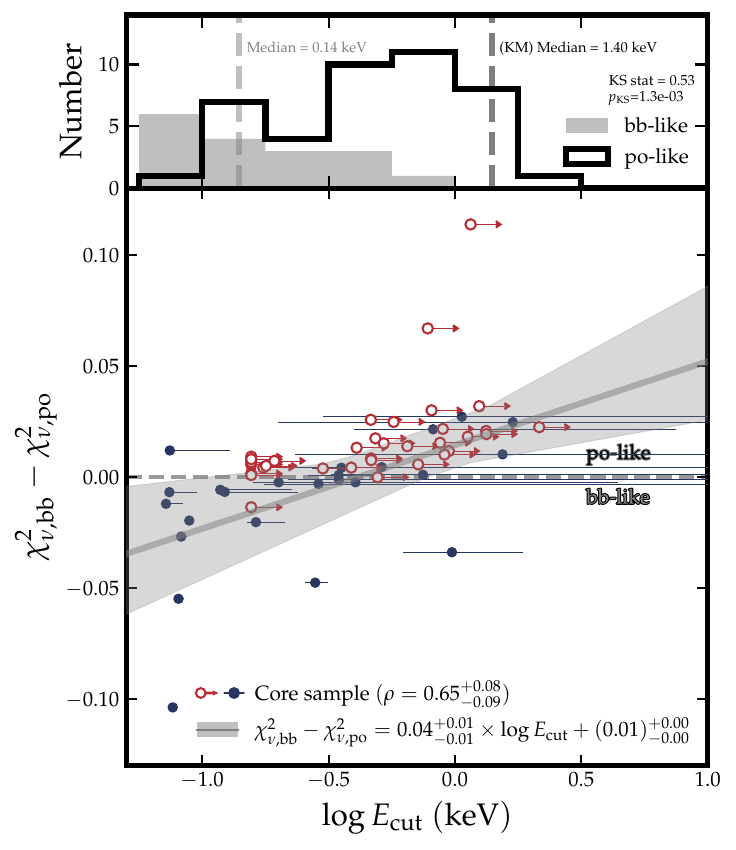}
    \caption{\textit{Upper panel}: The $E_\text{cut}$ distribution of bb-like sources and po-like sources. The Kaplan-Meier estimator is applied to estimate the median $E_\text{cut}$ for po-like sources. The KS test demonstrates a significant difference between the two distributions. \textit{Lower panel}: The correlation between the two parameters adopted to quantify soft excess profile: $E_\text{cut}$ and $\chi^2_{\nu,\text{bb}}-\chi^2_{\nu,\text{po}}$ (see text for definition). Sources with $E_\text{cut}$ measurements are denoted by blue filled circles, while those with lower limits are indicated by red circles (a convention that will be used throughout this paper). The Spearman's correlation coefficient $\rho$ is presented. The solid line represents the best-fit linear regression between the parameters (using the X-axis as the independent variable, and this convention will continue to be used for all linear regressions presented in this work), while the grey shaded region indicates the corresponding 1$\sigma$ confidence bands.}
    \label{fig:Ecvschi}
\end{figure}
\begin{deluxetable}{ccccc}
\tabletypesize{\scriptsize}
\tablecaption{The correlations between soft excess profile and broadband properties.\label{tab:ParTab1}}
\renewcommand{\arraystretch}{1.4}
\setlength{\tabcolsep}{3pt}
\tablehead{
\multirow{2}[0]{*}{Physical quantities} & 
\multicolumn{2}{c}{$\chi^2_{\nu,\text{bb}}-\chi^2_{\nu,\text{po}}$} & 
\multicolumn{2}{c}{$\log E_\text{cut}$}\\
\cmidrule(lr){2-3}
\cmidrule(lr){4-5}
\colhead{} &
\colhead{Spearman's $\rho$} & 
\colhead{p-value} & 
\colhead{Spearman's $\rho$} & 
\colhead{p-value}
}
\startdata
$\log q$\tablenotemark{1} & $0.57_{-0.09}^{+0.10}$ & $2.5e-06$ & $0.56_{-0.11}^{+0.09}$ & $2.3e-05$ \\ 
$\log L_\text{SE,0.5-2}/L_\text{UV}$\tablenotemark{2} & $0.36_{-0.11}^{+0.10}$ & $4.5e-03$ & $0.46_{-0.12}^{+0.11}$ & $5.2e-04$ \\ 
$\Gamma_\text{PC}$\tablenotemark{3} & $-0.36_{-0.11}^{+0.11}$ & $5.4e-03$ & $-0.46_{-0.09}^{+0.11}$ & $4.1e-04$ \\ 
$\log L_\text{UV}/L_\text{PC}$\tablenotemark{4} & $0.31_{-0.11}^{+0.11}$ & $0.02$ & $0.11_{-0.14}^{+0.11}$ & $0.37$ \\ 
$\log\lambda_\text{Edd}$\tablenotemark{5} & $0.17_{-0.14}^{+0.11}$ & $0.20$ & $-0.00_{-0.12}^{+0.13}$ & $0.53$ \\ 
\enddata
\tablenotetext{1}{The soft excess strength defined as Eq. \ref{q}.}
\tablenotetext{2}{The luminosity ratio of soft excess (0.5 -- 2 keV) to UV ($\nu_{2500\angstrom}L_{2500\angstrom}$).}
\tablenotetext{3}{The photon index of primary continuum (model 3).}
\tablenotetext{4}{The luminosity ratio of UV to primary continuum (0.5 -- 10 keV).}
\tablenotetext{5}{The Eddington ratio estimated with Eq. \ref{Edd}.}
\end{deluxetable}

In this work we adopt two parameters to quantify the soft excess spectral profile: 1) $E_\text{cut}$ of the soft excess from model 3; and 2) $\chi^2_{\nu,\text{bb}}-\chi^2_{\nu,\text{po}}$, i.e., the difference in the reduced chi-square when fitting soft excess with blackbody (model 1) or powerlaw (model 2). Note here we adopt the difference in reduced chi-square $\chi^2_{\nu}$, rather than the absolute difference in chi-square $\chi^2$, for further correlation analysis, as the latter depends on spectral quality, which varies significantly across our core sample (with degrees of freedom ranging from $\sim200$ to $\sim1800$). We plot in Fig. \ref{fig:Ecvschi} soft excess $E_\text{cut}$ versus $\chi^2_{\nu,\text{bb}}-\chi^2_{\nu,\text{po}}$, and quantify the correlation between them. We employ the \texttt{ASURV} package for quantitative analysis of censored data ($E_\text{cut}$). The generalized Spearman's rank $\rho$ implemented in \texttt{ASURV} is utilized to evaluate the strength of correlation, and the binned two-dimensional Kaplan-Meier method \citep{Schmitt1985} is employed for linear regression. Each correlation and regression parameter is provided with a $1\sigma$ confidence level, determined by bootstrapping points 400 times and recalculating these statistical quantities. In the case of linear regression, a shaded area indicative of $1\sigma$ confidence range is also depicted in the plot, with the bounds determined by the 400 bootstrapped regressions.

As shown in Fig. \ref{fig:Ecvschi}, we see a strong positive correlation between $E_\text{cut}$ of the soft excess and $\chi^2_{\nu,\text{bb}}-\chi^2_{\nu,\text{po}}$ (with Spearman's $\rho$ = $0.65^{+0.08}_{-0.09}$), that the soft excess profiles of sources with smaller/higher soft excess $E_\text{cut}$ are more likely bb-like/po-like. We also compare the $E_\text{cut}$ distribution between bb-like sources and po-like sources in the upper panel of Fig. \ref{fig:Ecvschi}, and find significant difference between the two groups (KS test null hypothesis probability $p_\text{KS}=1.3\times 10^{-3}$)\footnote{The KS test is performed using either the best fit $E_\text{cut}$ or the lower limit of $E_\text{cut}$ when it cannot be constrained. Consequently, the value of $1.3\times 10^{-3}$ should be considered an upper limit, and we anticipate an even more significant difference in $E_\text{cut}$ distribution between bb-like and po-like sources.}. Therefore these two parameters could similarly quantify the SE spectral profile.

We then assess the correlation between SE spectral profile and other physical properties. Specifically, we consider physical properties including the SE strength $\log q$, the ratio of SE luminosity to UV luminosity ($L_\text{SE}/L_\text{UV}$), the primary powerlaw continuum photon index ($\Gamma_\text{PC}$) obtained from model 3 fit, the ratio of UV luminosity to primary X-ray powerlaw continuum luminosity ($L_\text{UV}/L_\text{PC}$), as well as the Eddington ratio ($\lambda_\text{Edd}$). The UV luminosity is represented by $\nu_{2500\angstrom}L_{2500\angstrom}$, and the primary continuum luminosity is estimated from the 0.5 -- 10 keV band. The Spearman's $\rho$ values, corresponding $1\sigma$ uncertainty ranges, and p-values are summarized in Table \ref{tab:ParTab1}.

We see the strongest correlation lies between SE profile and the SE strength $\log q$ (see Table \ref{tab:ParTab1}). Such a correlation has already be qualitatively highlighted in \S\ref{Profile}. In this section we focus on quantitative analysis of this correlation. In \S\ref{cosp} we present and discuss the reliability of this strong correlation. A comparison of this correlation with the ionized-disk reflection model is detailed in \S \ref{relxilllp}. In \S \ref{relcpt} we propose a double-component scenario which could naturally account for the observed correlation. Finally we discuss the relation between SE profile and other physical properties in \S \ref{other}.

\subsection{The SE strength $\log q$ versus SE profile} \label{cosp}
\begin{figure*}
    \centering
    \includegraphics[width=1.8\columnwidth]{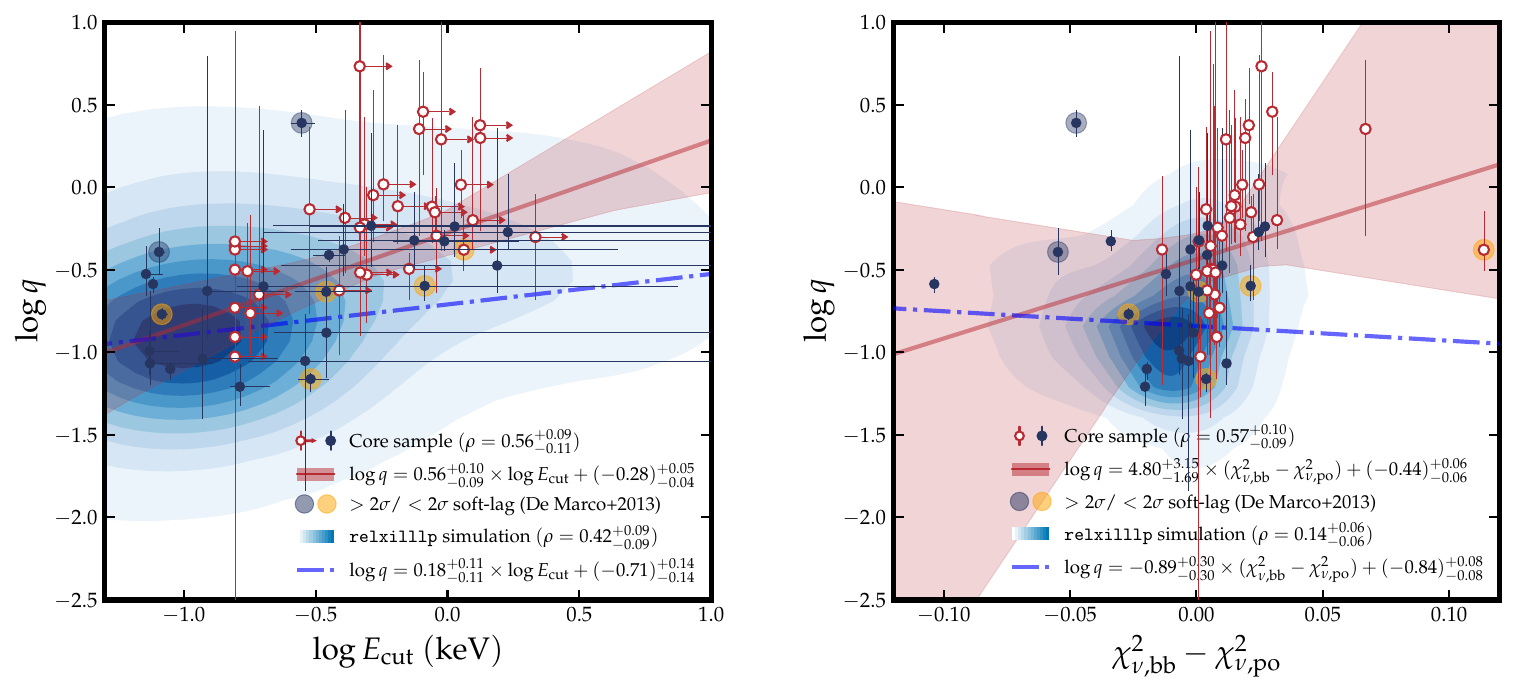}
    \caption{The SE strength $\log q$ versus spectral profile, with the latter quantified by SE cut-off energy $E_\text{cut}$ (left panel) and $\chi^2_{\nu,\text{bb}}-\chi^2_{\nu,\text{po}}$ (right panel). In each panel, the red solid line and shaded area plot the best-fit linear regression between parameters and the corresponding 1$\sigma$ confidence band. The \relxilllp~simulation results are overlaid as blue contour (KDE applied), along with the median linear fit relation (blue dashdot line). Sources with soft lag detection ($>2\sigma$) and non-detection ($<2\sigma$) from \citealt{DeMarco+2013} are marked with blue and yellow shaded circles respectively.}
    \label{fig:relxilllp}
\end{figure*}

\begin{figure}
    \centering
    \includegraphics[width=1\columnwidth]{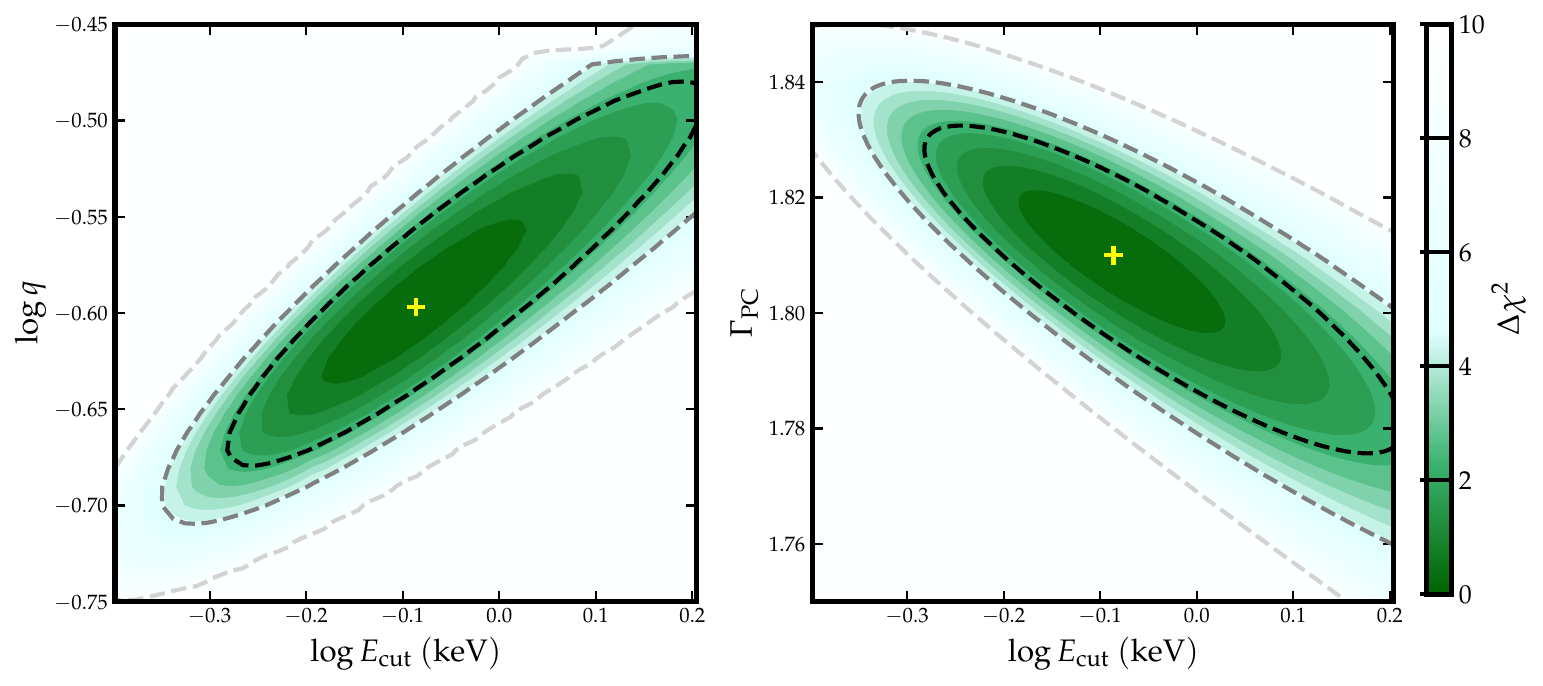}
    \caption{An example of \xspec~\texttt{steppar} results for $\log q\sim\log E_\text{cut}$ (left) and $\Gamma_\text{PC}\sim\log E_\text{cut}$ (right), based on the spectrum of MRK 110 (OBSID: 0852590201), to illustrate the degeneracy between parameters. The yellow plus denotes the best fit position, while the three dashed lines mark the $\Delta\chi^2=2.30, 4.61, 9.21$ confidence regions respectively. }
    \label{fig:degeneracy}
\end{figure}
In Fig. \ref{fig:relxilllp} we plot the correlation between SE strength $\log q$, and SE spectral profile (left panel: SE cut-off energy $E_\text{cut}$; right panel: $\chi^2_{\nu,\text{bb}}-\chi^2_{\nu,\text{po}}$). A linear relation between $\log q$ and $\log E_\text{cut}$ is observed:
\begin{equation}
    \log q = 0.56^{+0.10}_{-0.09}\times \log E_\text{cut} + (-0.28)^{+0.05}_{-0.04}\label{q_ecut_sp}
\end{equation}
The strong correlation, as estimated by Spearman's $\rho$, is $0.56^{+0.09}_{-0.11}$.

Regarding $\chi^2_{\nu,\text{bb}}-\chi^2_{\nu,\text{po}}$, we also observe a correlation $\rho$ as strong as $0.57^{+0.10}_{-0.09}$, and a linear relation:
\begin{equation}
    \log q = 4.80^{+3.15}_{-1.69}\times (\chi^2_{\nu,\text{bb}}-\chi^2_{\nu,\text{po}}) + (-0.44)^{+0.06}_{-0.06}\label{q_delchi_cp}
\end{equation}

We note that in fitting a spectrum with model 3, $\log E_\text{cut}$ and $\log q$ are degenerated (see the left panel of Fig. \ref{fig:degeneracy}). Such a degeneracy between the two parameters may yield artificial positive correlation between $\log q$ and $\log E_\text{cut}$ for a sample. We then perform simulations to testify this possibility. For each source in the core sample, we generate a fake spectrum based on the best-fit parameters of model 3 (using \texttt{fakeit} command in \xspec), and fit it again with model 3. For the artificial sample with 59 fake points in the $\log q$ -- $\log E_\text{cut}$ space, we re-compute the correlation. Due to the degeneracy between the two parameters (Fig. \ref{fig:degeneracy}), fitting a fake spectrum could possibly yield simultaneously larger (or smaller) $\log q$ and  $\log E_\text{cut}$ compared with the input model values, resulting in an excess correlation of the artificial sample compared with the observed $\rho=0.56$. However, by carrying out the above experiment for 10 times (each with different seeds when generating photon counts in \texttt{fakeit}), we find a mean value of $\rho$ = $0.63$ for the 10 correlation coefficients. The excess correlation ($0.63-0.56=0.07$) is only marginal considering the error bar of the observed $\rho$ ($0.56^{+0.09}_{-0.11}$), thereby demonstrating that the degeneracy cannot account for the strong correlation observed in the core sample.

On the other hand, the other spectral profile parameter $\chi^2_{\nu,\text{bb}}-\chi^2_{\nu,\text{po}}$ is determined from model 1 and 2 fit, and insusceptible to the potential degeneration with $\log q$ derived from model 3 fit. Therefore the strong correlation between $\log q$ and $\chi^2_{\nu,\text{bb}}-\chi^2_{\nu,\text{po}}$ also indicates a strong intrinsic correlation between SE strength and spectral profile.

\subsection{\relxilllp~simulation} \label{relxilllp}
To examine if the observed trend in Fig. \ref{fig:relxilllp} can be explained solely by the ionized disk reflection scenario for SE, we perform simulations using \relxilllp, a standard model predicting the relativistically-blurred reflection spectrum assuming a lamppost corona \citep{Dauser+2013,Dauser+2014,Garcia+2014,Dauser+2016a}. The simulated spectra are generated by \texttt{fakeit} command.

To find out the reasonable parameter ranges for \relxilllp, we first fit the core sample with the spectral model:
\begin{verbatim}
    phabs*zphabs*(relxilllp+zgauss)
\end{verbatim}
where for simplicity we fix $N_\text{H}$ of \verb|phabs| and \verb|zphabs| to the best-fit value from model 3, centroid energy and line width at 6.4 keV and 0.019 keV respectively for the narrow \feka~line (\texttt{zgauss}). The luminosity, coronal height \verb|h|, blackhole spin \verb|a|, inclination angle \verb|Incl|, primary continuum photon index \verb|gamma|, ionization parameter \verb|logxi|, iron abundance \verb|Afe|, and reflection fraction \verb|refl_frac| are free to vary. Based on the each set of best fit parameters (there are 59 sets in total), we generate 10 fake spectra, each with different seeds for Poisson realization. After that, we fit the 590 spectra with model 3, assuming the same parameter limits as those outlined in \S \ref{model}.

The simulation results are depicted as blue contours overlaid on Fig. \ref{fig:relxilllp}, employing Multivariate Kernel Density Estimation (KDE) implemented in \texttt{statsmodels} \citep{seabold2010statsmodels}. For each set of 59 simulated spectra (there are 10 sets in total), we compute the correlation coefficients and perform linear regressions. The mean values as well as standard deviations of these statistical quantities are provided in Fig. \ref{fig:relxilllp}.

The \relxilllp~simulation results generally do not align with those of the core sample. Firstly, the parameter space covered by \relxilllp~simulation differs from that of the core sample in Fig. \ref{fig:relxilllp}. A great portion ($\sim 70\%$) of the simulated spectra prefer a bb-like soft excess, with $\chi^2_{\nu,\text{bb}}-\chi^2_{\nu,\text{po}}<0$ and constrained $E_\text{cut}<0.3\ \text{keV}$. Note the quasi-blackbody profile for the SE component in the ionized reflection scenario arises from the blurred emission line structures at the soft band, but not due to a thermal process. Additionally, ionized disk reflection predicts only a mild correlation between $\log q$ and $\log E_\text{cut}$ ($\rho=0.42^{+0.09}_{-0.09}$), and a weak correlation between $\log q$ and $\chi^2_{\nu,\text{bb}}-\chi^2_{\nu,\text{po}}$ ($\rho=0.14^{+0.06}_{-0.06}$), which contrast the trend observed in the core sample. Moreover, the linear slopes of the simulation ($0.18^{+0.11}_{-0.11}$ for $\log q\sim\log E_\text{cut}$, and $-0.89^{+0.30}_{-0.30}$ for $\log q\sim\chi^2_{\nu,\text{bb}}-\chi^2_{\nu,\text{po}}$) do not match those of the core sample either (see Fig. \ref{fig:relxilllp}).

However, we note that the \relxilllp~simulation could reproduce the observed SE in some sources (i.e., those with bb-like SE profile or small $E_\text{cut}$, see also the overlap between the distribution peak of the \relxilllp~simulation and the core sample in Fig. \ref{fig:relxilllp}). Indeed, in the literature some individual sources here have been shown to favor the ionized reflection scenario. For example, \citealt{JiangJiang+2019} showed that the spectral variability of broad \feka~line in 1H 0419-577 ($\log q=-0.52,\ E_\text{cut}=0.07\ \text{keV},\ \chi^2_{\nu,\text{bb}}-\chi^2_{\nu,\text{po}}=-0.01$) can be explained by light-bending effect; \citealt{Garcia+2019} showed that the ionized reflection can produce the soft excess observed in MRK 509 ($\log q=-0.77,\ E_\text{cut}=0.08\ \text{keV},\ \chi^2_{\nu,\text{bb}}-\chi^2_{\nu,\text{po}}=-0.03$) in a more reasonable way compared with warm corona; and \citealt{Cackett+2013} argued a soft lag observed in ESO 113-10 ($\log q=-0.59,\ E_\text{cut}=0.08\ \text{keV},\ \chi^2_{\nu,\text{bb}}-\chi^2_{\nu,\text{po}}=-0.10$). Therefore, our findings support that part of bb-like soft excess could have disk reflection origin.

We also systematically examine the soft lag detection for the core sample. The soft lag is commonly interpreted as a signature of disk reflection \citep[e.g.][]{Emmanoulopoulos+2011,Kara+2013,Zoghbi+2015,Wilkins+2021,Bambi+2021,Hancock+2022}. For consistent measurement of soft lag, we match the core sample with the 32 sources in \citealt{DeMarco+2013} (choosing a larger sample from e.g. \citealt{Kara+2016} would result in a similar matched number). Out of the 7 overlapping sources, 2 showed soft lag detection ($>2\sigma$, PG 1211+143, RE J1034+396), represented by blue shaded circles in Fig. \ref{fig:relxilllp}, while the 5 sources with non-detection ($<2\sigma$, MRK 279, MRK 509, ESO 198-24, ESO 511-30, MRK 110, in decreasing soft lag significance order) are indicated with orange circles. Although the sample is small, we could see a trend that the two sources with soft lag detected tend to have bb-like SE profile and smaller $E_\text{cut}$, compared with the 5 sources with soft lag non-detected. This also suggests that the SE in AGNs with bb-like SE profile or smaller SE $E_\text{cut}$ are more likely dominated by ionized disk reflection, and such sources could be better candidates for X-ray reverberation mapping studies. 

In conclusion, while ionized disk reflection may explain the soft excess in some individual cases, it appears insufficient to consistently explain the soft excess profile and strength across the entire sample. 

\subsection{\relxilllp+\comptt~simulation} \label{relcpt}
\begin{figure*}
    \centering
    \includegraphics[width=1.8\columnwidth]{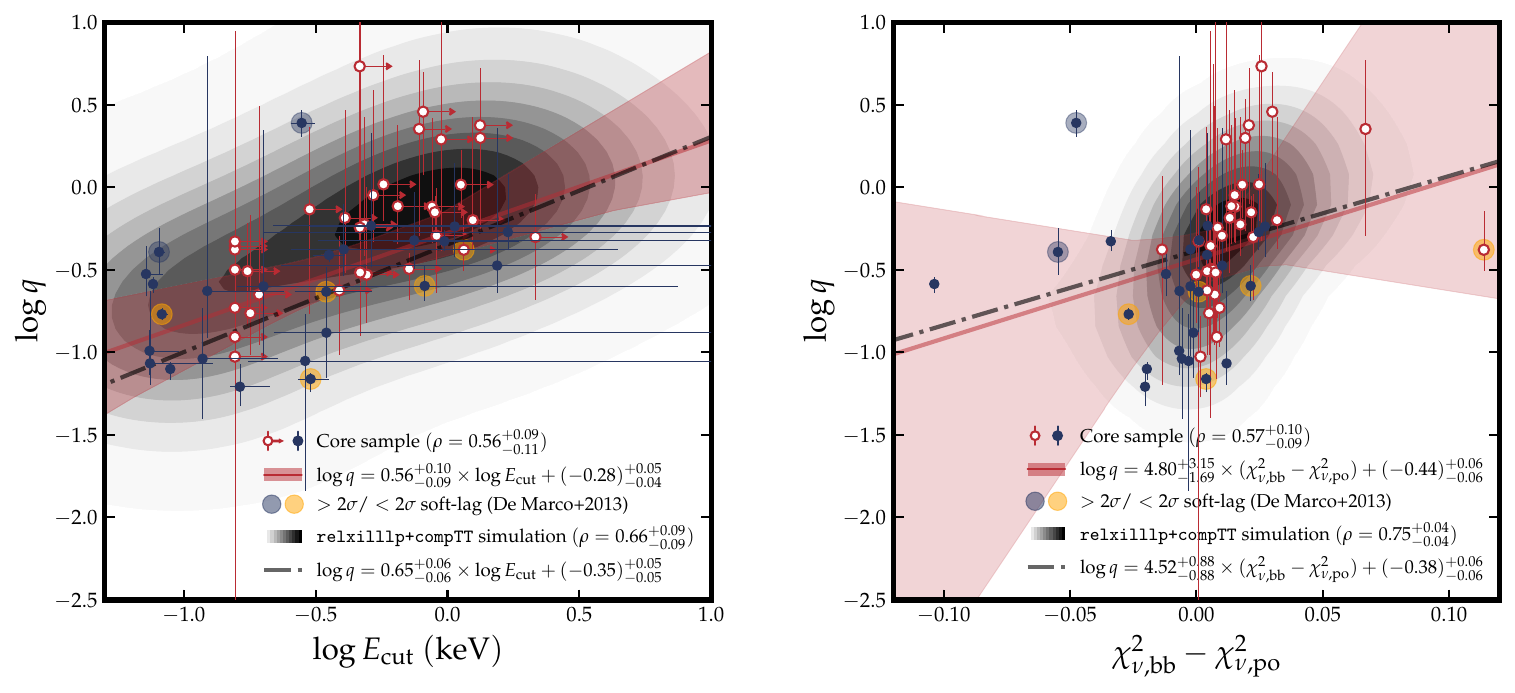}
    \caption{Similar to Fig. \ref{fig:relxilllp}, except that a hybrid input model (a mixture of ionized disk reflection and warm corona, i.e., \relxilllp+\comptt) is adopted for simulation. The KDE contours are shown in black, with a kernel bandwidth identical to that in Fig. \ref{fig:relxilllp}. The black dot-dashed lines represent the linear regression of the simulated data.}
    \label{fig:relcpt}
\end{figure*}

In \S \ref{relxilllp} we demonstrate that ionized disk reflection alone cannot account for the soft excess across the entire population. Conversely, if the soft excess is solely attributed to a warm corona (i.e. no reflection), it would be a struggle to explain the strong reflection feature as well as soft lags observed in some individual sources. Additionally, a pure warm corona predicts strong absorption \citep{Garcia+2019} in the soft band, contrary to the smooth spectrum observed in the core sample.

The observational challenges encountered when explaining soft excess with either ionized reflection or warm corona, have triggered theoretical studies in the literature to include for both. Studies have highlighted the pivotal roles played by illumination from the hot corona and internal heating from the disk in the formation of a warm corona, as well as generating a featureless soft excess observed in most type 1 AGNs \citep[e.g.,][]{Ballantyne2020,Petrucci+2020,Xiang+2022}. Therefore a warm corona accompanied by ionized reflection could be a natural case. 

In an attempt to \textit{self-consistently combine emission from a warm corona with relativistic reflection}, \citealt{Xiang+2022} introduced the \texttt{REXCOR} model, which quantifies the fraction of accretion energy dissipated in both hot corona ($f_X$) and warm corona ($h_f$). $f_X$ directly relates to the contribution from ionized reflection in the soft band. According to the theoretical predictions of \texttt{REXCOR} (see Fig. 4 of \citealt{Xiang+2022}), the SE spectral profile becomes more extended as $h_f$ increases. On the other hand, when $f_X$ becomes important, the ``cut-off'' position of the soft excess remains at a low value ($\lesssim 0.7\ \text{keV}$ from visual inspection). Based on this, a wide range of SE spectral profile can be produced when varying the relative contribution from the reflection and the warm corona. 

Therefore here we perform simulations to investigate whether a hybrid scenario (which includes both ionized disk reflection and warm corona emission) could reproduce the SE strength - profile correlation we have observed. The simulation setups and fitting procedures are similar to those in \S\ref{relxilllp}, with the addition of a warm corona component (\comptt) in the input model for the simulation:
\begin{verbatim}
    phabs*zphabs*(relxilllp+compTT+zgauss)
\end{verbatim}
The luminosity and optical depths (\verb|taup|) of \comptt, the luminosity, ionization parameter (\verb|logxi|) and photon index (\verb|gamma|) of \relxilllp, are obtained from fitting the core sample (59 different combinations in total). Because of the strong degeneracy between the ionized disk reflection component and the warm corona emission, properly decomposing the two components through spectra fitting could be challenging, and is deferred to a future work. Here we set up a toy hybrid scenario by simply fixing the warm corona plasma temperature \verb|kT| at 1 keV, seed photon temperature \verb|T0| at 5 eV, and the parameters of ionized disk reflection at their default values (\verb|h=6GM/c^2|, \verb|a=0.998|, \verb|Incl=30deg|, \verb|Afe=1|, \verb|refl_frac=1|). Similar to \S\ref{relxilllp}, a total of 590 artificial spectra are simulated, and fitted with model 3 to assess the SE strength as well as spectral profile. 

The comparison between the \relxilllp+\comptt~simulations and the core sample is demonstrated in Fig. \ref{fig:relcpt}. Remarkably, with an additional warm corona component, the simulations based on a toy hybrid model could nicely reproduce the observed distribution of SE strength and spectral profile, as well as the correlation between them. 

As a side note, we make some comments on the temperature of the warm corona. Traditionally, the warm corona is characterized by electron temperature ranging from 0.1 -- 1 keV, when fitting the observed soft excess with a single warm corona component \citep[e.g.,][]{Petrucci+2013,Petrucci+2018}. However, our work suggests that, when the contribution of ionized reflection is considered (with constant reflection fraction), even a warm corona with fixed temperature can produce a variety of soft excess profiles, with cut-off energy ranging from $\sim 0.1\ \text{keV}$ to $\sim 1\ \text{keV}$. This is in line with recent theoretical work by \citealt{Xiang+2022}, where they found that different ratios of warm corona/ionized reflection could lead to different soft excess shapes. This scheme is also consistent with the results of \cite{Boissay+2016} which found the soft excess shape varies with its strength while the hard X-ray reflection stays constant. Meanwhile we do acknowledge that in the real universe the warm corona temperature could vary from source to source, as a natural consequence of different physical conditions and heating -- cooling processes \citep[e.g.,][]{Rozanska2015,Petrucci+2020,Gronkiewicz+2023,Kawanaka&Mineshige2024}; however, to obtain an unbiased warm corona temperature (and other related quantities, e.g., optical depth) from spectral fitting, subtracting the contribution from ionized reflection is necessary. Such task generally requires higher spectral quality and could potentially be fulfilled with future mission, e.g., New Athena \citep[e.g.,][]{Nandra+2013}.

We finally note that the \texttt{REXCOR} model \citep{Xiang+2022}, which self-consistently combine X-ray emission from a warm corona with relativistic disk reflection, could also effectively reproduce the observed correlations in Fig. \ref{fig:relcpt}, similar to the simple  \relxilllp+\comptt~model. This finding further supports a hybrid nature for the soft excess in our sample. An extensive study leveraging this physically motivated model to analyze derived physical parameters and to compare \texttt{REXCOR} with other models, including \relxilllp+\comptt~ and phenomenological models, is deferred to a future dedicated work.

\subsection{SE spectral profile versus other broadband properties}\label{other}
Besides the strongest correlation between SE spectral profile and SE strength $\log q$, we also see mild correlation between SE spectral profile and other physical parameters. The mild correlation between SE profile and $L_\text{SE}/L_\text{UV}$ (with $\rho$ = 0.36 - 0.46) could be physically linked with that between SE profile and SE strength $\log q$, as while $\log q$ measures the strength of SE relative to the powerlaw component, $L_\text{SE}/L_\text{UV}$ measures the relative strength of SE relative to UV emission. Converting UV luminosity to X-ray powerlaw luminosity would bring in extra scatter, thus weaker correlation is seen between SE profile and $L_\text{SE}/L_\text{UV}$.

Similarly, we find a mild negative correlation between primary powerlaw continuum photon index $\Gamma_\text{PC}$ and SE profile, with $\rho=-0.36^{+0.11}_{-0.11}$ for $\chi^2_{\nu,\text{bb}}-\chi^2_{\nu,\text{po}}$ and $\rho=-0.46^{+0.11}_{-0.09}$ for $\log E_\text{cut}$. Noticing the degeneracy between $\log E_\text{cut}$ and $\Gamma_\text{PC}$ (see the right panel of Fig. \ref{fig:degeneracy}), we perform simulations following procedures similar to \S \ref{cosp}. The resulting correlation turns out to be similar to the original one within uncertainty range, suggesting that the anti-correlation between SE spectral profile and $\Gamma_\text{PC}$ should be real. The underlying physics however is unclear. As a mild anti-correlation between $\Gamma_\text{PC}$ and $\log q$ is also seen (Chen et al. in prep, Paper II), the observed correlation between $\Gamma_\text{PC}$ and SE profile could be a secondary effect. Meanwhile, a recent work \citep[][]{Kang&Wang2022} reports a positive correlation between hot corona temperature and primary continuum photon index. If the anti-correlation between $\Gamma_\text{PC}$ and SE profile we observed is intrinsic, this could suggest a negative correlation between hot corona temperature and warm corona temperature, indicating an interesting competing relation between the warm and hot corona. 

Meanwhile, it is unclear whether there is a correlation between $L_\text{UV}/L_\text{PC}$ and SE spectral profile. While we observe no correlation between $L_\text{UV}/L_\text{PC}$ and $\log E_\text{cut}$, a marginal correlation between $\chi^2_{\nu,\text{bb}}-\chi^2_{\nu,\text{po}}$ and $L_\text{UV}/L_\text{PC}$ is observed, with Spearman's $\rho = 0.31_{-0.11}^{+0.11}$ and a null hypothesis probability of $\sim2\%$. Nevertheless, the marginal correlation could align with our findings in \S\ref{relxilllp} and \S\ref{relcpt}, that if the warm corona emission is tightly correlated with UV emission (Chen et al. in prep., Paper II), higher $L_\text{UV}/L_\text{PC}$ would suggest higher relative contribution from the warm corona to soft excess compared with the ionized disk reflection, and thus making the SE spectral profile more po-like. 

 Finally, we do not find significant correlation between $\log\lambda_\text{Edd}$ and the SE spectral profile, however the relationship between the warm corona temperature and accretion rate is still uncertain, as the observed spectral profile depends not only on the warm corona temperature, but also the relative contribution of warm corona to ionized reflection. Notably, several recent studies have shown that the warm corona temperature exhibits a complex dependency on multiple physical parameters besides accretion rate, such as magnetic viscosity \citep{Gronkiewicz&Rozanska2020,Gronkiewicz+2023}. Additionally, the soft excess profile could be affected by relatively high seed photon temperatures \citep{Tang+2024}, which typically occur at high accretion rates. Studying the intriguing correlation between the warm corona temperature and these physical parameters requires future work that properly decomposes the ionized reflection and warm corona emission.

\section{Conclusions} \label{Conclusions}
In this first paper of our series, we build a sample of 59 unobscured type 1 AGNs with simultaneous X-ray and UV \xmm~observation  for extensive investigations of the soft X-ray excess in AGNs. In this paper we focus on the soft excess spectral profile, and the key conclusions drawn are as follows:

\begin{enumerate}
    \item We identify both po-like ($71\%$) and bb-like ($29\%$) soft excess profiles in the sample. A cut-off powerlaw can characterize both types of soft excesses in a uniform way, with typical cut-off energy $E_\text{cut}\sim 0.14\ \text{keV}$ for bb-like soft excess, and $\sim$ $1.40\ \text{keV}$ for po-like soft excess.

    \item We report for the first time the strong correlation between SE profile (characterized by soft excess $E_\text{cut}$, and $\chi^2_{\nu,\text{bb}}-\chi^2_{\nu,\text{po}}$ which is the difference in reduced chi-square when fitting soft excess with a blackbody or a powerlaw) and SE strength $\log q$ (parameterized as the luminosity ratio of soft excess to primary continuum in the 0.5 -- 2 keV band). The strong correlations, with $\rho=0.56^{+0.09}_{-0.11}$ between $\log q$ and $\log E_\text{cut}$, and $\rho=0.57^{+0.10}_{-0.09}$ between $\log q$ and $\chi^2_{\nu,\text{bb}}-\chi^2_{\nu,\text{po}}$, are confirmed from direct view of the ``unfolded'' soft excess spectra as well as quantitative assessments, and can not be attributed to parameter degeneracy. The correlations between SE profile and other physical parameters are much weaker or absent.

    \item We perform simulations showing that ionized disk reflection alone (simulated with \relxilllp) produces mostly bb-like soft excess, and can not recover the observed correlation between SE profile and strength, which is therefore insufficient to consistently explain the soft excess shape and strength across our sample.

    \item Remarkably, simulations assuming a toy hybrid model for the SE (i.e., a mixture of warm corona with temperature fixed at 1 keV and ionized reflection with all parameters except for ionization fixed at default values) can successfully reproduce the observed soft excess strength -- profile correlation. In this scheme, the observed SE profile is sensitive to the relative contribution of ionized disk reflection and warm corona emission. This underscores the importance of properly subtracting ionized disk reflection for unbiased measurement of warm corona temperature during spectral fitting.

\end{enumerate}

\section*{Acknowledgements}
This work was supported by the National Natural Science Foundation of China (grant nos. 12033006, 12192221, 12373016, 123B2042, 124B1007) and the Cyrus Chun Ying Tang Foundations. We are deeply grateful to Johannes Buchner from the Max-Planck Institute for Extraterrestrial Physics (MPE) for discussions on statistical methodologies, Shi-Fu Zhu and Jia-Lai Wang from the University of Science and Technology of China (USTC) for their insights into statistical analysis, Wen-Ke Ren from USTC for valuable suggestions on UV data analysis, and Fu-Guo Xie from the Shanghai Astronomical Observatory (SHAO) for discussions on the physical implications of our findings.

We also extend our heartfelt appreciation to the referee and the Scientific Editor for their thorough review and invaluable feedback, as well as to any other individuals whose contributions may not be explicitly mentioned here but whose advice and comments have significantly enriched this work.

The research utilized observations obtained with \xmm, an ESA science mission supported by contributions from ESA Member States and NASA. Additionally, we acknowledge the NASA/IPAC Extragalactic Database (NED), operated by the California Institute of Technology under contract with NASA \citep{NASA/IPACExtragalacticDatabaseNED2019}, for providing valuable data resources.

For bibliography, this research has made use of NASA's Astrophysics Data System, along with the adstex bibliography tool (\url{https://github.com/yymao/adstex}).

\software{
HEAsoft (v6.28; HEASARC 2014),
\xspec~\citep{Arnaud1996},
\asurv~\citep{Feigelson&Nelson1985,Isobe+1986},
\topcat~\citep{Taylor2005},
GNU Parallel Tool \citep{Tange2011a}
}

\appendix
\section{Potential effects of absorption on soft excess profile}\label{profile-abs}
\begin{figure}
    \centering
    \includegraphics[width=0.8\columnwidth]{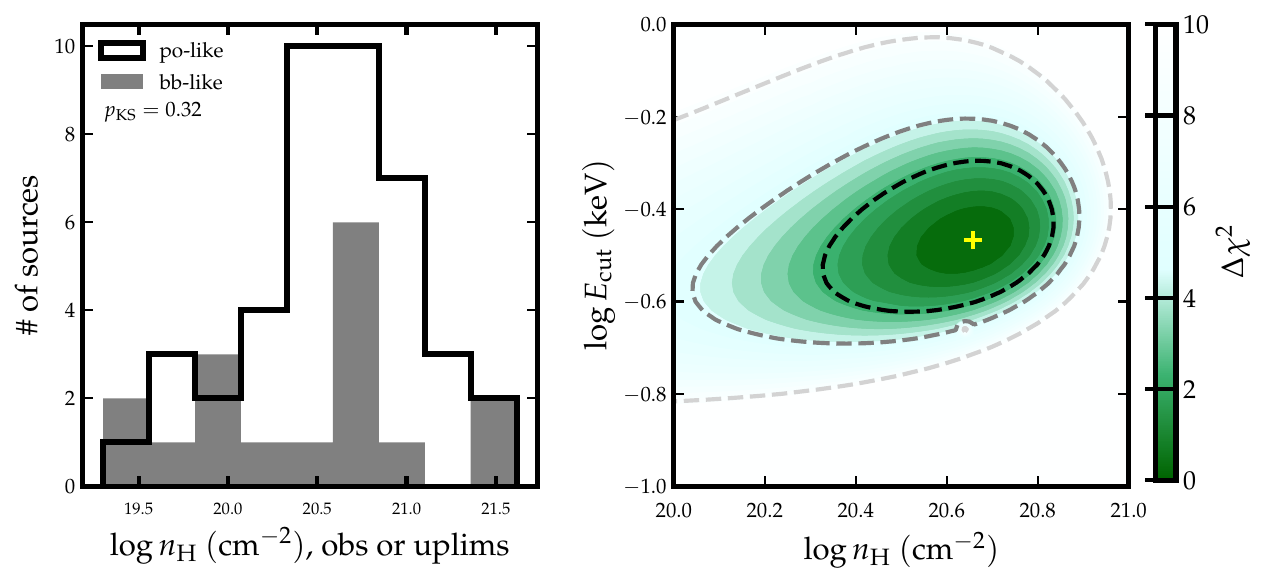}
    \caption{Left: The distribution of neutral column density $n_\mathrm{H}$ for bb-like and po-like sources in the core sample, derived from Model 3, considering either the best-fit values or upper limits. No significant difference is seen between the two subgroups. Right: An example of weak parameter degeneracy between the neutral column density $n_\mathrm{H}$ and $E_\mathrm{cut}$ for a representative source (ESO198-24, OBSID: 0305370101). }
    \label{fig:zphabs_bl_pl}
\end{figure}
\begin{figure}
    \centering
    \includegraphics[width=0.8\columnwidth]{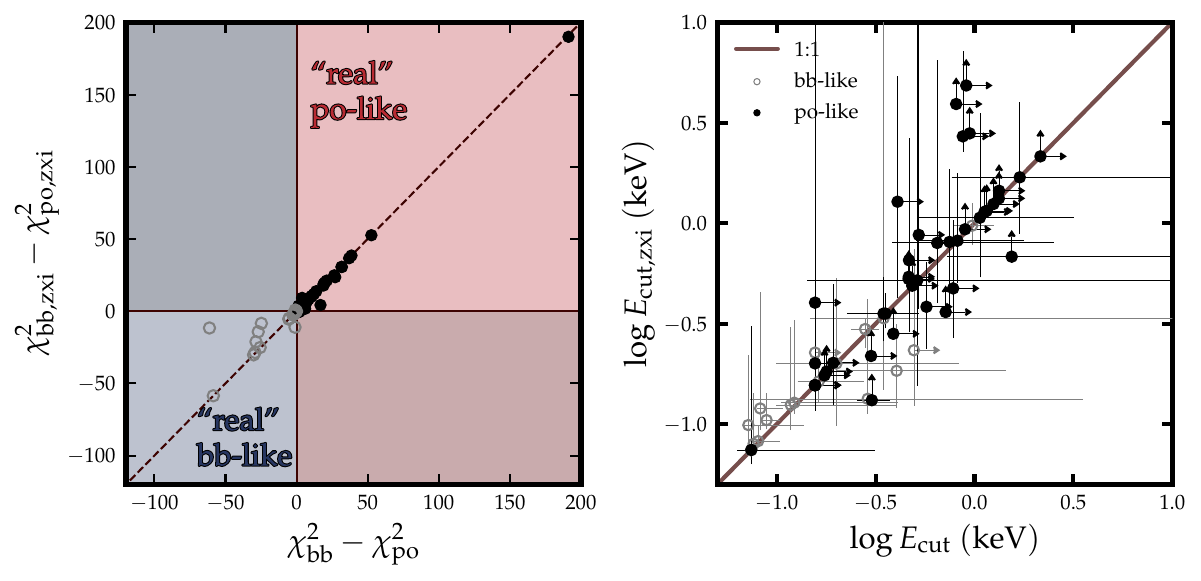}
    \caption{The comparison of soft excess spectral profile parameters (left: $\chi^2_\mathrm{bb}-\chi^2_\mathrm{po}$; right: $E_\mathrm{cut}$) before (X) / after (Y) adding an additional ionized absorber.}
    \label{fig:delchi2bbpo_ec_zxi2}
\end{figure}
In this work we identified both bb-like soft excess sources, whose spectral profiles are better fitted with blackbodies, and po-like sources, whose spectral profiles resemble powerlaws. However, it could be possible that some bb-like sources are in fact po-like, which suffer from strong absorption and thereby mimicking a bb-like characteristic (or vice versa). In this section we explore such possibilities and discuss whether such effects would influence our results.

The most common absorber in AGN is neutral absorption, which we have already taken into account in the spectral modeling (\phabs~in \S\ref{model}). By design, the bb-like sources are better fit by a blackbody than a power-law, assuming the presence of a neutral absorber with freely varying column density. Comparing the column density distributions (both best-fit values and upper limits) of bb-like and po-like sources (Fig. \ref{fig:zphabs_bl_pl}, left panel), we observe no significant differences. Additionally, for sources where both column density and cut-off energy can be constrained, we find only weak degeneracy between these parameters (right panel of Fig. \ref{fig:zphabs_bl_pl}). These indicate that neutral absorption plays a minimal role in determining the shape of the soft excess.

Ionized absorption, on the other hand, has the potential to significantly influence the soft excess profile. Monte Carlo simulations readily demonstrate that a bb-like profile can be mimicked by a po-like profile combined with ionized absorption. This is why we excluded spectra exhibiting strong ionized absorption when constructing the core sample in \S\ref{cosample}. Below, we further investigate whether the presence of weaker ionized absorption could impact the results of this study. We note that fitting the low-resolution PN spectra may lead to overestimated column densities of the ionized absorbers, due to the model's flexibility arising from parameter degeneracies. To that end, we first measure the ionized absorber column densities through fitting the corresponding high-resolution RGS spectra. For this analysis,  we only consider the first and most significant ionized absorber revealed by the RGS spectra, as it is expected to have the dominant impact on the soft excess shape compared to other absorbers. We then incorporate ionized absorption into model 1, 2 and 3 using the \zxipcf~component (assuming a fixed covering factor of 1). We impose an upper limit on the column density based on the RGS-derived value while allowing the ionization parameter to vary freely. With these modified models, we reassess the classification of sources as bb-like or po-like, and recalculate the soft excess cut-off energy. In the left panel of Fig. \ref{fig:delchi2bbpo_ec_zxi2} we compare $\chi^2_\mathrm{bb,zxi}-\chi^2_\mathrm{po,zxi}$ (the difference in chi-square between modified model 1 and 2) against $\chi^2_\mathrm{bb}-\chi^2_\mathrm{po}$. While incorporating ionized absorption improved the fit of the power law relative to the blackbody for a few bb-like sources, the majority retained the same sign for $\chi^2_\mathrm{bb}-\chi^2_\mathrm{po}$. This indicates that ionized absorption does not systematically alter the bb-like or po-like classification of our sources in the core sample. Consequently, the bb-like characteristic we identified is likely intrinsic rather than a distortion caused by ionized absorption. Moreover, as shown in the right panel of Fig. \ref{fig:delchi2bbpo_ec_zxi2}, the soft excess cut-off energy remains largely unchanged after accounting for the ionized absorption. Finally, incorporating the spectral fitting results that include ionized absorption does not affect the overall conclusions of this study.

\section{Table of Key Parameter values} \label{TabKeyPar}
In Table \ref{tab:ParTab2} we provide the fundamental physical parameters for each source in the core sample. The name of each source is listed in column 2, followed by the corresponding \xmm~observation ID in column 3. Columns 4 and 5 document the Galactic column density (in units of $10^{20}\ \text{cm}^{-2}$) and redshift, respectively. Column 6 and 7 display the black hole mass, along with the corresponding reference. The Eddington ratio, shown in column 8, is estimated using the formula:
\begin{equation}
\lambda_\text{Edd}=\frac{L_\text{bol}}{L_\text{Edd}}=\frac{\text{BC}_{2500\angstrom}\times (\nu_{2500\angstrom}L_{2500\angstrom})}{1.5\times10^{38}\times M_\text{BH}}\label{Edd}
\end{equation}
assuming a bolometric correction factor $\text{BC}_{2500\angstrom}=2.75$ \citep{Krawczyk+2013a}. Finally we present the UV luminosity ($L_\text{UV}\equiv\nu_{2500\angstrom}L_{2500\angstrom}$, see \S \ref{om}), as well as $\alpha_\text{oX}$ in column 9 and 10 respectively.

In Table \ref{tab:ParTab3} we present the spectral fitting results for model 1, 2 and 3 (see \S \ref{model}). For the primary continuum (\pexrav), we list its luminosity ($L_\text{PC,bb},L_\text{PC,po},L_\text{PC,cpl}$) in logarithmic scale (0.5 -- 10 keV), photon index ($\Gamma_\text{PC,bb},\Gamma_\text{PC,po},\Gamma_\text{PC,cpl}$), and reflection fraction ($R_\text{bb},R_\text{po},R_\text{cpl}$). For the soft excess component, the luminosity ($L_\text{SE,0.5-2,bb},L_\text{SE,0.5-2,po},L_\text{SE,0.5-2,cpl}$) in logarithmic scale, ``shape'' parameter ($T_\text{bb}$ for model 1,$\Gamma_\text{SE,po}$ for model 2, and $E_\text{cut}, \Gamma_\text{SE,cpl}$ for model 3), and soft excess strength ($\log q$, see Eq. \ref{q}) under model 3, are provided. The reduced chi-squares for each model, indicative of fitting statistics, are also presented  ($\chi^2_{\nu,\text{bb}},\chi^2_{\nu,\text{po}},\chi^2_{\nu,\text{cpl}}$).

In X-ray spectral fitting, the luminosity of primary continuum and soft excess within 0.5 -- 2 keV band are degenerated with each other. Therefore to estimate the uncertainty of soft excess strength correctly, we multiply soft excess with a \texttt{const} component, and link the soft excess luminosity (0.5 -- 2 keV) with primary continuum luminosity (0.5 -- 2 keV) to the same value. The error bars of $q$ are estimated from the uncertainties of \texttt{const} given by \texttt{error} command in \xspec.

We note that for some bb-like sources ($\chi^2_{\nu,\text{bb}}-\chi^2_{\nu,\text{po}}<0$, and $E_\text{cut}\lesssim 0.2\ \text{keV}$), the soft excess photon indices when fitting with model 3 ($\texttt{cutoffpl}$) are poorly constrained or even unphysical ($\lesssim 1$). But this is understandable. Firstly, when $E_\text{cut}\lesssim 0.2\ \text{keV}$, the powerlaw part of the cut-off powerlaw falls outside the effective \pn~bandpass. Secondly, as mentioned in \S \ref{relxilllp}, the bb-like profile possibly indicates an ionized-reflection origin, where the soft excess is mainly contributed by blurred emission lines rather than a Comptonized process. Therefore the cut-off powerlaw here only serves as a phenomenological shape function, and the photon index does not contain physical meanings.

We also note the median best-fit value for the reflection fraction $R$ is 0.76 for model 2 (powerlaw) and 1.34 for model 3 (cut-off powerlaw). For model 1 (blackbody), a much stronger reflection fraction is generally required, with a significant portion of sources having only a lower limit on $R$. Such excessively strong reflection values are generally unphysical, highlighting the inadequacy of a blackbody model in characterizing a broad powerlaw-like soft excess (e.g., see Fig. \ref{fig:eg_polike}).

\startlongtable
\begin{deluxetable*}{cccccccccc}
\tabletypesize{\scriptsize}
\renewcommand{\arraystretch}{0.95}
\tablecaption{The fundamental physical parameters of sources in the core sample.\label{tab:ParTab2}}
\tablehead{
\colhead{ID} & 
\colhead{Name} & 
\colhead{OBSID} & 
\colhead{$N_\text{H,gal}$ (10$^{20}\text{cm}^{-2}$)} & 
\colhead{z} & 
\colhead{$\log M_\text{BH}(M_\odot)$} & 
\colhead{Ref} & 
\colhead{$\log\lambda_\text{Edd}$} & 
\colhead{$\log L_\text{UV}\ (\text{erg}\ \text{s}^{-1})$} & 
\colhead{$\alpha_\text{oX}$}
}
\startdata
1 & 1H0419-577 & 0148000601 & 1.16 & 0.10 & 8.34 & 4 & -1.10 & $44.98$ & $1.32$ \\
2 & 3C382 & 0790600301 & 6.19 & 0.06 & 8.01 & 4 & -1.38 & $44.36$ & $1.05$ \\
3 & 3C390.3 & 0203720201 & 3.67 & 0.06 & 8.64 & 4 & -2.21 & $44.17$ & $1.00$ \\
4 & ESO113-10 & 0301890101 & 1.97 & 0.03 & 6.71 & 3 & -1.38 & $43.07$ & $1.21$ \\
5 & ESO141-55 & 0913190101 & 4.94 & 0.04 & 7.99 & 4 & -1.12 & $44.60$ & $1.33$ \\
6 & ESO198-24 & 0305370101 & 2.68 & 0.05 & 8.50 & 4 & -2.46 & $43.78$ & $1.15$ \\
7 & ESO244-17 & 0103860901 & 1.54 & 0.02 & 6.78 & 5 & -1.63 & $42.89$ & $1.21$ \\
8 & ESO359-19 & 0201130101 & 0.61 & 0.06 & 8.57 & 4 & -2.79 & $43.52$ & $1.20$ \\
9 & ESO438-9 & 0903040401 & 4.80 & 0.02 & 6.87 & 4 & -1.15 & $43.46$ & $1.36$ \\
10 & ESO511-30 & 0502090201 & 4.34 & 0.02 & 7.23 & 4 & -1.34 & $43.62$ & $1.19$ \\
11 & ESO548-81 & 0312190601 & 2.37 & 0.01 & 7.96 & 4 & -2.41 & $43.28$ & $1.25$ \\
12 & Fairall9 & 0605800401 & 2.86 & 0.05 & 8.30 & 4 & -1.70 & $44.34$ & $1.29$ \\
13 & HE1029-1401 & 0890410101 & 5.73 & 0.09 & 9.08 & 1 & -1.87 & $44.95$ & $1.34$ \\
14 & HE1143-1810 & 0795580101 & 3.08 & 0.03 & 7.39 & 4 & -1.16 & $43.96$ & $1.21$ \\
15 & HE2254-3712 & 0205390101 & 1.02 & 0.04 & 6.58 & 1 & -0.93 & $43.38$ & $1.29$ \\
16 & LBQS1228+1116 & 0306630201 & 2.32 & 0.24 & 8.70 & 6 & -1.47 & $44.96$ & $1.33$ \\
17 & MCG+04-22-042 & 0312191401 & 3.60 & 0.03 & 7.27 & 4 & -1.14 & $43.86$ & $1.18$ \\
18 & MRK110 & 0852590201 & 1.27 & 0.04 & 7.29 & 4 & -1.21 & $43.81$ & $1.10$ \\
19 & MRK1148 & 0801890301 & 4.02 & 0.06 & 7.75 & 4 & -1.21 & $44.28$ & $1.11$ \\
20 & MRK1383 & 0852210101 & 2.56 & 0.09 & 9.01 & 4 & -1.81 & $44.94$ & $1.42$ \\
21 & MRK279 & 0302480401 & 1.28 & 0.03 & 7.43 & 4 & -1.17 & $44.00$ & $1.19$ \\
22 & MRK352 & 0312190101 & 5.31 & 0.01 & 7.56 & 4 & -2.26 & $43.03$ & $1.22$ \\
23 & MRK359 & 0830551101 & 4.37 & 0.02 & 6.05 & 4 & -0.89 & $42.89$ & $1.35$ \\
24 & MRK493 & 0744290101 & 1.96 & 0.03 & 6.17 & 1 & -0.54 & $43.37$ & $1.37$ \\
25 & MRK50 & 0601781001 & 1.71 & 0.02 & 7.42 & 4 & -1.90 & $43.25$ & $1.08$ \\
26 & MRK509 & 0130720201 & 3.93 & 0.03 & 8.05 & 4 & -1.29 & $44.49$ & $1.28$ \\
27 & MRK590 & 0201020201 & 2.77 & 0.03 & 7.57 & 4 & -2.27 & $43.03$ & $1.14$ \\
28 & MRK705 & 0783270401 & 3.43 & 0.03 & 7.08 & 4 & -1.03 & $43.79$ & $1.20$ \\
29 & MRK732 & 0601780201 & 2.07 & 0.03 & 6.59 & 4 & -1.11 & $43.21$ & $1.27$ \\
30 & MRK926 & 0790640101 & 2.87 & 0.05 & 7.98 & 4 & -1.48 & $44.23$ & $1.04$ \\
31 & NGC1566 & 0820530401 & 0.71 & 0.00 & 6.83 & 4 & -2.29 & $42.27$ & $1.25$ \\
32 & NGC4748 & 0723100401 & 3.56 & 0.01 & 6.41 & 4 & -1.01 & $43.13$ & $1.25$ \\
33 & NGC7213 & 0111810101 & 1.08 & 0.00 & 7.13 & 4 & -3.04 & $41.82$ & $1.01$ \\
34 & PG0052+251 & 0841480101 & 3.96 & 0.15 & 8.46 & 4 & -1.33 & $44.86$ & $1.18$ \\
35 & PG0804+761 & 0605110101 & 3.34 & 0.10 & 8.24 & 1 & -0.62 & $45.35$ & $1.42$ \\
36 & PG0947+396 & 0841482301 & 1.70 & 0.21 & 8.68 & 1 & -1.61 & $44.80$ & $1.32$ \\
37 & PG0953+414 & 0111290201 & 1.09 & 0.23 & 8.24 & 1 & -0.48 & $45.50$ & $1.35$ \\
38 & PG1048+342 & 0109080701 & 1.72 & 0.17 & 8.37 & 1 & -1.46 & $44.64$ & $1.34$ \\
39 & PG1115+407 & 0111290301 & 1.41 & 0.15 & 7.67 & 1 & -0.49 & $44.91$ & $1.40$ \\
40 & PG1116+215 & 0201940101 & 1.21 & 0.18 & 8.53 & 1 & -0.64 & $45.62$ & $1.50$ \\
41 & PG1149-110 & 0783271401 & 3.20 & 0.05 & 7.68 & 4 & -1.98 & $43.43$ & $1.11$ \\
42 & PG1202+282 & 0109080101 & 1.74 & 0.17 & 8.10 & 4 & -1.07 & $44.77$ & $1.25$ \\
43 & PG1211+143 & 0745110301 & 2.63 & 0.08 & 7.49 & 1 & -0.53 & $44.70$ & $1.43$ \\
44 & PG1216+069 & 0111291101 & 1.51 & 0.33 & 9.20 & 1 & -1.30 & $45.63$ & $1.47$ \\
45 & PG1307+085 & 0841481401 & 2.10 & 0.15 & 8.72 & 6 & -1.32 & $45.13$ & $1.37$ \\
46 & PG1402+261 & 0830470101 & 1.22 & 0.16 & 7.94 & 1 & -0.73 & $44.95$ & $1.36$ \\
47 & PG1415+451 & 0109080501 & 0.74 & 0.11 & 8.01 & 1 & -1.36 & $44.38$ & $1.38$ \\
48 & PG1427+480 & 0109080901 & 1.61 & 0.22 & 8.29 & 6 & -1.18 & $44.84$ & $1.30$ \\
49 & PG1440+356 & 0005010201 & 0.90 & 0.08 & 7.47 & 1 & -0.58 & $44.63$ & $1.42$ \\
50 & PG1545+210 & 0783272101 & 3.46 & 0.26 & 8.88 & 4 & -1.38 & $45.24$ & $1.23$ \\
51 & PG2304+042 & 0783272701 & 5.44 & 0.04 & 8.09 & 4 & -1.91 & $43.92$ & $1.13$ \\
52 & PKS0558-504 & 0555170301 & 3.28 & 0.14 & 8.48 & 2 & -1.10 & $45.11$ & $1.19$ \\
53 & PKS2135-14 & 0092850201 & 4.15 & 0.20 & 9.10 & 4 & -2.18 & $44.66$ & $1.07$ \\
54 & Q1821+643 & 0506210701 & 3.50 & 0.30 & 9.23 & 4 & -0.74 & $46.23$ & $1.34$ \\
55 & RBS1756 & 0201130301 & 3.29 & 0.03 & 7.84 & 4 & -2.38 & $43.19$ & $1.14$ \\
56 & REJ1034+396 & 0865011501 & 1.25 & 0.04 & 6.81 & 1 & -1.22 & $43.33$ & $1.30$ \\
57 & SBS1301+540 & 0312192001 & 1.73 & 0.03 & 7.67 & 4 & -1.97 & $43.44$ & $1.08$ \\
58 & WISEAJ144414.66+0633 & 0841480701 & 2.57 & 0.21 & 8.45 & 6 & -1.37 & $44.81$ & $1.20$ \\
59 & ZW229-015 & 0672530301 & 5.34 & 0.03 & 6.91 & 4 & -1.71 & $42.94$ & $1.14$ \\
\enddata
\tablecomments{
Column (1): source ID in the core sample. 
Column (2): common name of source. 
Column (3): \xmm~observation ID. 
Column (4): galactic absorption in units of 10$^{20}\text{cm}^{-2}$. 
Column (5): redshift. 
Column (6)-(7): blackhole mass and references. 1: \citealt{Bianchi+2009}, 2: \citealt{GV2012}, 3: \citealt{Busch+2014}, 4: \citealt{Koss+2017}, 5: \citealt{Williams+2018}, 6: \citealt{ChenChen+2022}. 
Column (8): Eddington ratio estimated with \ref{Edd}. 
Column (9): UV luminosity ($2500L_{2500\angstrom}$). 
Column (10): $\alpha_\text{oX}$. 
}
\end{deluxetable*}

\startlongtable
\begin{longrotatetable}
\movetabledown=0.8in
\begin{deluxetable*}{ccccccccccccccccccccc}
\tabletypesize{\tiny}
\tablecaption{The spectral fitting results of sources in the core sample. \label{tab:ParTab3}}
\tablewidth{0pt}
\renewcommand{\arraystretch}{1.5}
\setlength{\tabcolsep}{1pt}
\tablehead{ 
 & 
\multicolumn{6}{c}{\texttt{Model 1 (\texttt{blackbody})}} & 
\multicolumn{6}{c}{\texttt{Model 2 (\texttt{powerlaw})}} & 
\multicolumn{8}{c}{\texttt{Model 3 (\texttt{cutoffpl})}} \\
\cmidrule(lr){2-7}
\cmidrule(lr){8-13}
\cmidrule(lr){14-21} 
\colhead{ID} & 
\colhead{$\log L_\text{PC,bb}$} & 
\colhead{$\Gamma_\text{PC,bb}$} & 
\colhead{$R_\text{bb}$} & 
\colhead{$\log L_\text{SE,0.5-2,bb}$} & 
\colhead{$T_\text{bb}$} & 
\colhead{$\chi^2_{\nu,\text{bb}}$} & 
\colhead{$\log L_\text{PC,po}$} & 
\colhead{$\Gamma_\text{PC,po}$} & 
\colhead{$R_\text{po}$} & 
\colhead{$\log L_\text{SE,0.5-2,po}$} & 
\colhead{$\Gamma_\text{SE,po}$} & 
\colhead{$\chi^2_{\nu,\text{po}}$} & 
\colhead{$\log L_\text{PC,cpl}$} & 
\colhead{$\Gamma_\text{PC,cpl}$} & 
\colhead{$R_\text{cpl}$} & 
\colhead{$\log L_\text{SE,0.5-2,cpl}$} & 
\colhead{$E_\text{cut}$} & 
\colhead{$\Gamma_\text{SE,cpl}$} & 
\colhead{$\log q$} & 
\colhead{$\chi^2_{\nu,\text{cpl}}$} \\
\colhead{} & 
\colhead{(erg$\ $s$^{-1}$)} & 
\colhead{} & 
\colhead{} & 
\colhead{(erg$\ $s$^{-1}$)} & 
\colhead{(keV)} & 
\colhead{} & 
\colhead{(erg$\ $s$^{-1}$)} & 
\colhead{} & 
\colhead{} & 
\colhead{(erg$\ $s$^{-1}$)} & 
\colhead{} & 
\colhead{} & 
\colhead{(erg$\ $s$^{-1}$)} & 
\colhead{} & 
\colhead{} & 
\colhead{(erg$\ $s$^{-1}$)} & 
\colhead{(keV)} & 
\colhead{} & 
\colhead{} & 
\colhead{} 
}
\startdata
1 & 44.61$_{-0.01}^{+0.01}$ & 1.820$_{-0.039}^{+0.043}$ & 3.64$_{-0.74}^{+0.35}$ & 43.65$_{-0.15}^{+0.14}$ & 0.08$_{-0.00}^{+0.00}$ & 1.013 & 44.61$_{-0.01}^{+0.01}$ & 1.833$_{-0.039}^{+0.041}$ & 3.72$_{-0.6}^{+0.51}$ & 44.05$_{-0.25}^{+0.20}$ & 6.11$_{-0.26}^{+0.25}$ & 1.025 & 44.61$_{-0.01}^{+0.01}$ & 1.824$_{-0.036}^{+0.037}$ & 3.61$_{-0.55}^{+0.29}$ & 43.50$_{-0.03}^{+0.02}$ & 0.07$_{-0.0}^{+0.07}$ & <-0.72$_{}^{}$ & -0.52$_{-0.13}^{+0.17}$ & 1.006 \\ 
2 & 44.71$_{-0.00}^{+0.00}$ & 1.904$_{-0.008}^{+0.039}$ & 2.85$_{-0.21}^{+0.13}$ & 43.53$_{-0.13}^{+0.18}$ & 0.10$_{-0.00}^{+0.00}$ & 1.029 & 44.67$_{-0.01}^{+0.01}$ & 1.732$_{-0.038}^{+0.030}$ & 0.48$_{-0.26}^{+0.58}$ & 43.52$_{-0.12}^{+0.28}$ & 3.52$_{-0.36}^{+0.39}$ & 1.048 & 44.70$_{-0.00}^{+0.00}$ & 1.862$_{-0.018}^{+0.011}$ & 2.30$_{-0.38}^{+0.26}$ & 43.06$_{-0.05}^{+0.05}$ & 0.09$_{-0.0}^{+0.02}$ & <-1.74$_{}^{}$ & -1.10$_{-0.07}^{+0.04}$ & 1.035 \\ 
3 & 44.66$_{-0.00}^{+0.00}$ & 1.832$_{-0.013}^{+0.014}$ & 1.90$_{-0.37}^{+0.32}$ & 42.93$_{-0.36}^{+0.21}$ & 0.12$_{-0.01}^{+0.01}$ & 1.001 & 44.60$_{-0.47}^{+0.02}$ & 1.689$_{-0.172}^{+0.046}$ & 0.78$_{-0.78}^{+0.47}$ & 43.60$_{-0.23}^{+0.51}$ & 2.66$_{-0.66}^{+0.54}$ & 1.004 & 44.65$_{-0.04}^{+0.00}$ & 1.786$_{-0.129}^{+0.017}$ & 0.53$_{-0.53}^{+0.6}$ & 43.17$_{-0.76}^{+0.28}$ & 0.29$_{-0.22}^{+3.27}$ & <2.67$_{}^{}$ & -1.05$_{-0.79}^{+0.28}$ & 1.003 \\ 
4 & 42.98$_{-0.00}^{+0.00}$ & 2.230$_{-0.071}^{+0.059}$ & 3.99$_{-1.0}^{+0.8}$ & 42.42$_{-0.19}^{+0.14}$ & 0.08$_{-0.00}^{+0.00}$ & 1.171 & 42.94$_{-0.01}^{+0.01}$ & 2.050$_{-0.057}^{+0.049}$ & <0.20$_{}^{}$ & 42.17$_{-0.06}^{+0.06}$ & 5.02$_{-0.33}^{+0.35}$ & 1.275 & 42.96$_{-0.01}^{+0.01}$ & 2.136$_{-0.034}^{+0.033}$ & 2.62$_{-1.51}^{+1.41}$ & 42.03$_{-0.04}^{+0.03}$ & 0.08$_{-0.0}^{+0.01}$ & <-2.42$_{}^{}$ & -0.59$_{-0.06}^{+0.04}$ & 1.177 \\ 
5 & 44.24$_{-0.00}^{+0.00}$ & 2.193$_{-0.006}^{+0.007}$ & 4.91$_{-0.21}^{+4.91}$ & 42.75$_{-0.04}^{+0.03}$ & 0.12$_{-0.00}^{+0.00}$ & 1.074 & 44.21$_{-0.01}^{+0.01}$ & 2.129$_{-0.021}^{+0.021}$ & 4.77$_{-0.19}^{+0.21}$ & 43.18$_{-0.10}^{+0.09}$ & 3.39$_{-0.16}^{+0.23}$ & 1.108 & 44.17$_{-0.01}^{+0.01}$ & 1.862$_{-0.030}^{+0.026}$ & 0.01$_{-0.01}^{+0.28}$ & 43.46$_{-0.04}^{+0.04}$ & 0.97$_{-0.19}^{+0.28}$ & 2.38$_{-0.15}^{+0.13}$ & -0.33$_{-0.06}^{+0.07}$ & 1.052 \\ 
6 & 43.91$_{-0.00}^{+0.00}$ & 1.851$_{-0.012}^{+0.018}$ & 2.39$_{-0.5}^{+0.31}$ & 42.35$_{-0.25}^{+0.18}$ & 0.13$_{-0.01}^{+0.01}$ & 1.054 & 43.83$_{-0.04}^{+0.02}$ & 1.588$_{-0.065}^{+0.039}$ & 0.07$_{-0.07}^{+1.22}$ & 43.17$_{-0.06}^{+0.06}$ & 3.05$_{-0.40}^{+0.32}$ & 1.054 & 43.90$_{-0.00}^{+0.00}$ & 1.815$_{-0.015}^{+0.016}$ & 1.63$_{-1.63}^{+0.44}$ & 42.84$_{-0.25}^{+0.15}$ & 0.35$_{-0.12}^{+0.17}$ & 1.83$_{-1.27}^{+0.68}$ & -0.63$_{-0.24}^{+0.15}$ & 1.045 \\ 
7 & 42.81$_{-0.01}^{+0.01}$ & 1.987$_{-0.044}^{+0.046}$ & >0.84$_{}^{}$ & 41.24$_{-0.34}^{+0.31}$ & 0.13$_{-0.04}^{+0.02}$ & 0.954 & 42.62$_{-0.49}^{+0.27}$ & 1.581$_{-1.490}^{+0.890}$ & 0.01$_{-0.01}^{+2.62}$ & 42.23$_{-0.39}^{+0.19}$ & 2.59$_{-1.49}^{+0.86}$ & 0.946 & 42.78$_{-0.08}^{+0.01}$ & 1.826$_{-0.376}^{+0.085}$ & 0.38$_{-0.38}^{+2.17}$ & 41.73$_{-0.26}^{+0.60}$ & >0.19$_{}^{}$ & <-1.89$_{}^{}$ & -0.65$_{-0.31}^{+1.15}$ & 0.945 \\ 
8 & 43.54$_{-0.01}^{+0.01}$ & 1.938$_{-0.068}^{+0.073}$ & 4.96$_{-4.11}^{+4.96}$ & 42.52$_{-0.63}^{+0.34}$ & 0.11$_{-0.01}^{+0.02}$ & 0.931 & 43.48$_{-0.32}^{+0.03}$ & 1.603$_{-0.457}^{+0.102}$ & 0.64$_{-0.64}^{+4.2}$ & 43.02$_{-0.45}^{+0.33}$ & 3.94$_{-1.85}^{+0.94}$ & 0.937 & 43.52$_{-0.01}^{+0.01}$ & 1.803$_{-0.050}^{+0.034}$ & 2.94$_{-2.27}^{+2.52}$ & 41.92$_{-0.72}^{+0.51}$ & 0.12$_{-0.02}^{+0.29}$ & <2.62$_{}^{}$ & -1.04$_{-0.37}^{+0.31}$ & 0.936 \\ 
9 & 43.03$_{-0.00}^{+0.00}$ & 2.044$_{-0.045}^{+0.055}$ & 4.98$_{-3.0}^{+4.98}$ & 41.74$_{-0.13}^{+0.26}$ & 0.09$_{-0.01}^{+0.01}$ & 1.149 & 43.01$_{-0.00}^{+0.00}$ & 1.981$_{-0.156}^{+0.079}$ & 3.87$_{-3.66}^{+3.87}$ & 41.97$_{-0.24}^{+0.33}$ & 4.46$_{-1.15}^{+1.41}$ & 1.156 & 43.03$_{-0.01}^{+0.01}$ & 2.061$_{-0.050}^{+0.049}$ & 4.95$_{-3.5}^{+4.95}$ & 41.71$_{-0.14}^{+0.11}$ & 0.07$_{-0.01}^{+0.11}$ & <1.00$_{}^{}$ & -0.99$_{-0.15}^{+0.13}$ & 1.149 \\ 
10 & 43.62$_{-0.00}^{+0.00}$ & 2.007$_{-0.000}^{+0.000}$ & 1.36$_{-0.19}^{+0.96}$ & 42.38$_{-0.02}^{+0.02}$ & 0.12$_{-0.00}^{+0.00}$ & 1.154 & 43.53$_{-0.01}^{+0.01}$ & 1.766$_{-0.027}^{+0.018}$ & 1.64$_{-0.33}^{+0.65}$ & 42.96$_{-0.06}^{+0.03}$ & 2.99$_{-0.06}^{+0.13}$ & 1.040 & 43.58$_{-0.04}^{+0.02}$ & 1.857$_{-0.038}^{+0.041}$ & 1.76$_{-0.67}^{+0.35}$ & 42.80$_{-0.09}^{+0.13}$ & >1.15$_{}^{}$ & 3.01$_{-0.2}^{+0.12}$ & -0.38$_{-0.13}^{+0.23}$ & 1.040 \\ 
11 & 43.12$_{-0.01}^{+0.01}$ & 1.993$_{-0.033}^{+0.037}$ & 3.14$_{-1.39}^{+1.47}$ & 41.71$_{-0.13}^{+0.23}$ & 0.09$_{-0.01}^{+0.01}$ & 0.905 & 43.11$_{-0.02}^{+0.01}$ & 1.925$_{-0.075}^{+0.062}$ & 2.41$_{-1.2}^{+0.66}$ & 41.96$_{-0.22}^{+0.29}$ & 4.40$_{-0.81}^{+1.25}$ & 0.893 & 43.12$_{-0.02}^{+0.01}$ & 2.001$_{-0.096}^{+0.033}$ & 3.12$_{-1.18}^{+1.56}$ & 41.69$_{-0.12}^{+0.38}$ & 0.07$_{-0.01}^{+0.24}$ & <5.66$_{}^{}$ & -1.07$_{-0.13}^{+0.44}$ & 0.894 \\ 
12 & 44.09$_{-0.00}^{+0.00}$ & 2.019$_{-0.006}^{+0.009}$ & 4.75$_{-0.32}^{+4.75}$ & 42.32$_{-0.06}^{+0.19}$ & 0.11$_{-0.01}^{+0.01}$ & 1.055 & 43.89$_{-0.13}^{+0.06}$ & 1.509$_{-0.158}^{+0.076}$ & 0.03$_{-0.03}^{+0.81}$ & 43.52$_{-0.05}^{+0.08}$ & 2.63$_{-0.26}^{+0.26}$ & 1.023 & 44.00$_{-0.24}^{+0.02}$ & 1.635$_{-0.256}^{+0.060}$ & 0.01$_{-0.01}^{+0.68}$ & 43.33$_{-0.12}^{+0.27}$ & >1.25$_{}^{}$ & 2.54$_{-0.25}^{+0.32}$ & -0.20$_{-0.17}^{+0.63}$ & 1.024 \\ 
13 & 44.55$_{-0.00}^{+0.00}$ & 2.138$_{-0.008}^{+0.013}$ & >4.80$_{}^{}$ & 43.28$_{-0.04}^{+0.04}$ & 0.12$_{-0.00}^{+0.00}$ & 1.095 & 44.38$_{-0.08}^{+0.04}$ & 1.591$_{-0.112}^{+0.073}$ & 0.01$_{-0.01}^{+0.78}$ & 44.11$_{-0.02}^{+0.02}$ & 3.10$_{-0.30}^{+0.31}$ & 1.077 & 44.44$_{-0.16}^{+0.03}$ & 1.684$_{-0.187}^{+0.069}$ & 0.04$_{-0.04}^{+0.89}$ & 44.01$_{-0.13}^{+0.11}$ & >1.12$_{}^{}$ & 2.91$_{-0.48}^{+0.47}$ & 0.02$_{-0.20}^{+0.21}$ & 1.078 \\ 
14 & 43.93$_{-0.00}^{+0.00}$ & 2.051$_{-0.011}^{+0.012}$ & >4.95$_{}^{}$ & 42.63$_{-0.04}^{+0.04}$ & 0.11$_{-0.01}^{+0.01}$ & 1.065 & 43.80$_{-0.03}^{+0.03}$ & 1.627$_{-0.081}^{+0.058}$ & 1.08$_{-0.52}^{+0.33}$ & 43.32$_{-0.06}^{+0.05}$ & 2.97$_{-0.11}^{+0.20}$ & 1.040 & 43.87$_{-0.09}^{+0.02}$ & 1.750$_{-0.152}^{+0.048}$ & 1.26$_{-0.63}^{+0.3}$ & 43.15$_{-0.08}^{+0.09}$ & 1.69$_{-0.93}^{+13.8}$ & 2.85$_{-0.29}^{+0.26}$ & -0.27$_{-0.11}^{+0.35}$ & 1.039 \\ 
15 & 43.15$_{-0.01}^{+0.01}$ & 2.031$_{-0.039}^{+0.041}$ & >4.23$_{}^{}$ & 42.09$_{-0.08}^{+0.07}$ & 0.12$_{-0.01}^{+0.01}$ & 1.144 & 42.92$_{-0.15}^{+0.14}$ & 1.500$_{-0.328}^{+0.272}$ & 0.00$_{-0.0}^{+3.71}$ & 42.75$_{-0.12}^{+0.08}$ & 2.81$_{-0.22}^{+0.78}$ & 1.077 & 42.93$_{-0.16}^{+1.01}$ & 1.522$_{-0.348}^{+0.360}$ & 0.00$_{-0.0}^{+3.52}$ & 42.74$_{-0.33}^{+0.08}$ & >0.78$_{}^{}$ & 2.85$_{-0.27}^{+0.74}$ & 0.35$_{-0.65}^{+0.42}$ & 1.079 \\ 
16 & 44.59$_{-0.00}^{+0.00}$ & 2.300$_{-0.029}^{+0.028}$ & >4.44$_{}^{}$ & 43.26$_{-0.10}^{+0.14}$ & 0.10$_{-0.02}^{+0.02}$ & 1.096 & 44.41$_{-0.14}^{+0.06}$ & 1.818$_{-0.223}^{+0.119}$ & 0.36$_{-0.36}^{+1.28}$ & 44.08$_{-0.12}^{+0.13}$ & 3.11$_{-0.30}^{+0.30}$ & 1.080 & 44.48$_{-0.20}^{+0.04}$ & 1.887$_{-0.283}^{+0.108}$ & 0.35$_{-0.35}^{+1.83}$ & 43.98$_{-0.14}^{+0.22}$ & >0.87$_{}^{}$ & 3.01$_{-0.49}^{+0.39}$ & -0.12$_{-0.22}^{+0.54}$ & 1.082 \\ 
17 & 43.89$_{-0.01}^{+0.01}$ & 2.081$_{-0.037}^{+0.033}$ & 3.66$_{-1.79}^{+3.64}$ & 42.14$_{-0.35}^{+0.39}$ & 0.09$_{-0.04}^{+0.04}$ & 0.922 & 43.45$_{-0.27}^{+0.28}$ & 1.204$_{-0.592}^{+0.443}$ & >0.00$_{}^{}$ & 42.78$_{-0.60}^{+0.49}$ & 2.36$_{-0.14}^{+0.35}$ & 0.911 & 43.67$_{-0.48}^{+0.16}$ & 1.422$_{-0.798}^{+0.358}$ & <1.80$_{}^{}$ & 43.39$_{-0.36}^{+0.12}$ & >0.95$_{}^{}$ & 2.39$_{-0.31}^{+0.35}$ & 0.29$_{-0.68}^{+1.07}$ & 0.912 \\ 
18 & 44.05$_{-0.00}^{+0.00}$ & 2.002$_{-0.008}^{+0.008}$ & 4.12$_{-0.23}^{+0.28}$ & 42.45$_{-0.05}^{+0.04}$ & 0.11$_{-0.01}^{+0.01}$ & 1.040 & 43.94$_{-0.02}^{+0.03}$ & 1.662$_{-0.070}^{+0.050}$ & 0.84$_{-0.48}^{+0.23}$ & 43.34$_{-0.07}^{+0.09}$ & 2.78$_{-0.15}^{+0.27}$ & 1.019 & 44.02$_{-0.01}^{+0.01}$ & 1.807$_{-0.037}^{+0.024}$ & 1.05$_{-0.25}^{+0.32}$ & 43.01$_{-0.07}^{+0.10}$ & 0.82$_{-0.31}^{+0.96}$ & 2.41$_{-0.47}^{+0.36}$ & -0.60$_{-0.09}^{+0.13}$ & 1.013 \\ 
19 & 44.50$_{-0.01}^{+0.01}$ & 1.933$_{-0.037}^{+0.038}$ & 2.40$_{-1.42}^{+2.4}$ & 43.39$_{-0.09}^{+0.10}$ & 0.14$_{-0.01}^{+0.01}$ & 0.956 & 44.29$_{-0.16}^{+0.12}$ & 1.428$_{-0.336}^{+0.211}$ & <1.49$_{}^{}$ & 44.03$_{-0.08}^{+0.08}$ & 2.72$_{-1.20}^{+0.88}$ & 0.958 & 44.47$_{-0.04}^{+0.02}$ & 1.744$_{-0.132}^{+0.070}$ & 0.01$_{-0.01}^{+3.01}$ & 43.66$_{-0.12}^{+0.17}$ & 0.40$_{-0.2}^{+1.04}$ & 1.25$_{-1.81}^{+1.22}$ & -0.38$_{-0.16}^{+0.27}$ & 0.951 \\ 
20 & 44.30$_{-0.00}^{+0.00}$ & 2.104$_{-0.012}^{+0.012}$ & >4.36$_{}^{}$ & 43.18$_{-0.03}^{+0.04}$ & 0.10$_{-0.00}^{+0.00}$ & 0.948 & 44.21$_{-0.02}^{+0.01}$ & 1.718$_{-0.044}^{+0.040}$ & 0.64$_{-0.64}^{+1.1}$ & 43.69$_{-0.04}^{+0.04}$ & 3.42$_{-0.12}^{+0.13}$ & 0.927 & 44.23$_{-0.04}^{+0.03}$ & 1.752$_{-0.074}^{+0.072}$ & 0.65$_{-0.65}^{+0.76}$ & 43.64$_{-0.10}^{+0.08}$ & >0.90$_{}^{}$ & 3.35$_{-0.44}^{+0.19}$ & -0.15$_{-0.15}^{+0.10}$ & 0.927 \\ 
21 & 44.01$_{-0.00}^{+0.00}$ & 2.086$_{-0.010}^{+0.012}$ & 3.87$_{-0.11}^{+0.26}$ & 42.37$_{-0.11}^{+0.18}$ & 0.09$_{-0.01}^{+0.01}$ & 0.984 & 44.00$_{-0.00}^{+0.00}$ & 2.028$_{-0.020}^{+0.011}$ & 2.83$_{-0.27}^{+0.14}$ & 42.61$_{-0.07}^{+0.09}$ & 4.50$_{-0.38}^{+0.38}$ & 0.980 & 44.01$_{-0.01}^{+0.00}$ & 2.040$_{-0.017}^{+0.010}$ & 2.81$_{-0.18}^{+0.16}$ & 42.50$_{-0.08}^{+0.06}$ & 0.30$_{-0.05}^{+0.07}$ & 2.65$_{-0.47}^{+1.84}$ & -1.16$_{-0.08}^{+0.04}$ & 0.979 \\ 
22 & 42.97$_{-0.01}^{+0.01}$ & 1.876$_{-0.021}^{+0.023}$ & >3.58$_{}^{}$ & 41.66$_{-0.06}^{+0.13}$ & 0.09$_{-0.01}^{+0.01}$ & 0.972 & 42.94$_{-0.01}^{+0.01}$ & 1.703$_{-0.055}^{+0.045}$ & 2.55$_{-2.55}^{+2.55}$ & 41.99$_{-0.07}^{+0.09}$ & 3.95$_{-0.37}^{+0.42}$ & 0.967 & 42.94$_{-0.02}^{+0.01}$ & 1.715$_{-0.030}^{+0.059}$ & 2.66$_{-2.66}^{+2.66}$ & 41.97$_{-0.15}^{+0.09}$ & >0.72$_{}^{}$ & 3.86$_{-1.0}^{+0.49}$ & -0.49$_{-0.19}^{+0.10}$ & 0.968 \\ 
23 & 42.48$_{-0.01}^{+0.01}$ & 2.007$_{-0.032}^{+0.033}$ & >4.76$_{}^{}$ & 41.24$_{-0.09}^{+0.10}$ & 0.13$_{-0.01}^{+0.01}$ & 1.011 & 42.23$_{-0.08}^{+0.13}$ & 1.324$_{-0.183}^{+0.228}$ & 1.40$_{-0.79}^{+0.4}$ & 42.02$_{-0.10}^{+0.05}$ & 2.61$_{-0.11}^{+0.53}$ & 0.981 & 42.22$_{-0.08}^{+0.20}$ & 1.290$_{-0.202}^{+0.172}$ & 0.92$_{-0.79}^{+0.65}$ & 42.02$_{-0.30}^{+0.03}$ & >0.81$_{}^{}$ & 2.61$_{-0.39}^{+0.16}$ & 0.46$_{-0.39}^{+0.24}$ & 0.981 \\ 
24 & 42.94$_{-0.01}^{+0.01}$ & 2.340$_{-0.037}^{+0.042}$ & >4.76$_{}^{}$ & 41.80$_{-0.12}^{+0.09}$ & 0.13$_{-0.01}^{+0.01}$ & 0.946 & 42.59$_{-0.20}^{+0.12}$ & 1.405$_{-0.395}^{+0.238}$ & <1.55$_{}^{}$ & 42.69$_{-0.04}^{+0.03}$ & 3.00$_{-0.32}^{+0.39}$ & 0.921 & 42.82$_{-0.30}^{+0.04}$ & 1.854$_{-0.541}^{+0.132}$ & 0.66$_{-0.66}^{+1.17}$ & 42.45$_{-0.14}^{+0.23}$ & >0.57$_{}^{}$ & 2.49$_{-0.57}^{+0.72}$ & 0.02$_{-0.22}^{+0.79}$ & 0.919 \\ 
25 & 43.48$_{-0.00}^{+0.00}$ & 2.131$_{-0.085}^{+0.074}$ & >4.01$_{}^{}$ & 41.95$_{-0.49}^{+0.48}$ & 0.11$_{-0.05}^{+0.03}$ & 1.032 & 43.34$_{-0.00}^{+0.00}$ & 1.869$_{-0.732}^{+1.990}$ & 3.88$_{-2.76}^{+3.87}$ & 42.86$_{-0.98}^{+0.34}$ & 2.77$_{-2.33}^{+2.25}$ & 1.026 & 43.43$_{-0.44}^{+0.06}$ & 1.935$_{-0.815}^{+0.228}$ & 3.74$_{-2.4}^{+3.74}$ & 42.70$_{-0.93}^{+0.49}$ & >0.16$_{}^{}$ & 2.87$_{-1.63}^{+2.07}$ & -0.36$_{-1.04}^{+1.30}$ & 1.029 \\ 
26 & 44.25$_{-0.00}^{+0.00}$ & 1.848$_{-0.015}^{+0.015}$ & 2.01$_{-0.68}^{+0.73}$ & 43.07$_{-0.03}^{+0.06}$ & 0.09$_{-0.00}^{+0.00}$ & 0.873 & 44.24$_{-0.01}^{+0.01}$ & 1.797$_{-0.026}^{+0.023}$ & 0.72$_{-0.72}^{+0.47}$ & 43.22$_{-0.05}^{+0.05}$ & 4.55$_{-0.28}^{+0.30}$ & 0.900 & 44.24$_{-0.00}^{+0.00}$ & 1.830$_{-0.019}^{+0.019}$ & 2.07$_{-0.75}^{+0.78}$ & 43.08$_{-0.03}^{+0.03}$ & 0.08$_{-0.0}^{+0.03}$ & <-1.23$_{}^{}$ & -0.77$_{-0.04}^{+0.04}$ & 0.884 \\ 
27 & 43.19$_{-0.01}^{+0.01}$ & 1.765$_{-0.026}^{+0.025}$ & 2.42$_{-0.43}^{+0.38}$ & 41.62$_{-0.17}^{+0.15}$ & 0.16$_{-0.02}^{+0.02}$ & 0.949 & 42.96$_{-0.31}^{+0.17}$ & 1.314$_{-0.482}^{+0.239}$ & 0.82$_{-0.66}^{+2.14}$ & 42.06$_{-0.54}^{+0.54}$ & 2.25$_{-2.25}^{+0.68}$ & 0.950 & 43.19$_{-0.03}^{+0.01}$ & 1.714$_{-0.079}^{+0.037}$ & 1.82$_{-0.69}^{+0.91}$ & 41.84$_{-0.20}^{+0.30}$ & 0.35$_{-0.2}^{+249.86}$ & 0.67$_{-2.9}^{+1.65}$ & -0.88$_{-0.27}^{+0.36}$ & 0.947 \\ 
28 & 43.70$_{-0.00}^{+0.00}$ & 2.281$_{-0.028}^{+0.029}$ & >4.09$_{}^{}$ & 42.53$_{-0.08}^{+0.07}$ & 0.12$_{-0.01}^{+0.01}$ & 0.925 & 43.33$_{-0.08}^{+0.08}$ & 1.385$_{-0.178}^{+0.164}$ & 0.11$_{-0.11}^{+0.11}$ & 43.38$_{-0.04}^{+0.03}$ & 2.77$_{-0.08}^{+0.10}$ & 0.924 & 43.65$_{-0.04}^{+0.02}$ & 2.035$_{-0.125}^{+0.073}$ & 2.54$_{-2.54}^{+1.39}$ & 42.98$_{-0.16}^{+0.18}$ & 0.75$_{-0.36}^{+1.77}$ & 2.39$_{-0.78}^{+0.5}$ & -0.32$_{-0.17}^{+0.29}$ & 0.919 \\ 
29 & 42.96$_{-0.01}^{+0.01}$ & 1.695$_{-0.055}^{+0.054}$ & <2.08$_{}^{}$ & 41.61$_{-0.27}^{+0.17}$ & 0.19$_{-0.03}^{+0.02}$ & 1.000 & 42.21$_{-0.70}^{+0.74}$ & 0.945$_{-1.675}^{+2.056}$ & 0.01$_{-0.01}^{+1.44}$ & 42.53$_{-2.76}^{+0.03}$ & 1.91$_{-0.09}^{+1.00}$ & 1.014 & 42.96$_{-0.08}^{+0.02}$ & 1.690$_{-0.201}^{+0.081}$ & 0.01$_{-0.01}^{+1.84}$ & 41.65$_{-0.33}^{+0.29}$ & >0.16$_{}^{}$ & 1.79$_{-3.35}^{+1.69}$ & -0.38$_{-0.82}^{+0.44}$ & 1.002 \\ 
30 & 44.64$_{-0.00}^{+0.00}$ & 1.886$_{-0.012}^{+0.014}$ & 2.46$_{-0.43}^{+0.46}$ & 42.98$_{-0.10}^{+0.12}$ & 0.13$_{-0.01}^{+0.01}$ & 0.984 & 44.48$_{-0.22}^{+0.06}$ & 1.562$_{-0.259}^{+0.094}$ & 0.76$_{-0.76}^{+1.07}$ & 43.96$_{-0.17}^{+0.18}$ & 2.51$_{-0.44}^{+0.60}$ & 0.973 & 44.60$_{-0.26}^{+0.01}$ & 1.693$_{-0.288}^{+0.037}$ & <1.61$_{}^{}$ & 43.66$_{-0.20}^{+0.33}$ & 1.54$_{-0.82}^{+32.83}$ & 2.44$_{-0.45}^{+0.65}$ & -0.47$_{-0.17}^{+0.83}$ & 0.974 \\ 
31 & 42.11$_{-0.00}^{+0.00}$ & 1.913$_{-0.011}^{+0.011}$ & >4.96$_{}^{}$ & 40.27$_{-0.30}^{+0.20}$ & 0.09$_{-0.02}^{+0.01}$ & 1.086 & 42.09$_{-0.03}^{+0.01}$ & 1.792$_{-0.019}^{+0.031}$ & 2.80$_{-0.49}^{+0.3}$ & 40.62$_{-0.15}^{+0.10}$ & 3.33$_{-0.68}^{+0.45}$ & 1.064 & 42.02$_{-0.09}^{+0.05}$ & 1.595$_{-0.165}^{+0.095}$ & 2.10$_{-0.34}^{+0.4}$ & 41.22$_{-0.41}^{+0.18}$ & >2.15$_{}^{}$ & 2.44$_{-0.22}^{+0.26}$ & -0.30$_{-0.38}^{+0.26}$ & 1.064 \\ 
32 & 42.98$_{-0.01}^{+0.01}$ & 2.441$_{-0.035}^{+0.052}$ & 4.78$_{-1.47}^{+4.78}$ & 41.72$_{-0.15}^{+0.29}$ & 0.10$_{-0.01}^{+0.01}$ & 0.965 & 42.79$_{-0.17}^{+0.08}$ & 1.935$_{-0.283}^{+0.177}$ & <1.38$_{}^{}$ & 42.58$_{-0.16}^{+0.09}$ & 3.39$_{-0.45}^{+0.45}$ & 0.952 & 42.90$_{-0.25}^{+0.04}$ & 2.097$_{-0.404}^{+0.151}$ & 0.01$_{-0.01}^{+1.03}$ & 42.41$_{-0.26}^{+0.25}$ & >0.41$_{}^{}$ & 3.18$_{-0.9}^{+0.62}$ & -0.18$_{-0.33}^{+0.65}$ & 0.953 \\ 
33 & 42.24$_{-0.00}^{+0.00}$ & 1.775$_{-0.011}^{+0.014}$ & 0.65$_{-0.4}^{+0.42}$ & 40.62$_{-0.08}^{+0.13}$ & 0.16$_{-0.02}^{+0.01}$ & 1.021 & 41.97$_{-0.29}^{+0.17}$ & 1.459$_{-0.294}^{+0.149}$ & 0.26$_{-0.26}^{+1.59}$ & 41.35$_{-0.64}^{+0.27}$ & 2.16$_{-2.16}^{+0.32}$ & 1.041 & 42.24$_{-0.00}^{+0.00}$ & 1.777$_{-0.016}^{+0.013}$ & 0.68$_{-0.53}^{+0.34}$ & 40.61$_{-0.11}^{+0.13}$ & 0.16$_{-0.04}^{+0.11}$ & <-1.00$_{}^{}$ & -1.21$_{-0.12}^{+0.13}$ & 1.022 \\ 
34 & 44.86$_{-0.00}^{+0.00}$ & 2.119$_{-0.016}^{+0.017}$ & >3.70$_{}^{}$ & 43.37$_{-0.11}^{+0.09}$ & 0.14$_{-0.02}^{+0.01}$ & 0.976 & 44.60$_{-0.22}^{+0.17}$ & 1.487$_{-0.159}^{+0.229}$ & 0.96$_{-0.96}^{+0.8}$ & 44.44$_{-0.08}^{+0.05}$ & 2.69$_{-0.30}^{+0.76}$ & 0.955 & 44.62$_{-0.25}^{+0.16}$ & 1.481$_{-0.372}^{+1.481}$ & <2.30$_{}^{}$ & 44.44$_{-0.35}^{+0.06}$ & >1.33$_{}^{}$ & 2.77$_{-0.54}^{+0.62}$ & 0.38$_{-0.48}^{+0.35}$ & 0.956 \\ 
35 & 44.73$_{-0.00}^{+0.00}$ & 2.386$_{-0.019}^{+0.028}$ & >4.56$_{}^{}$ & 43.67$_{-0.06}^{+0.05}$ & 0.12$_{-0.01}^{+0.01}$ & 0.922 & 44.55$_{-0.12}^{+0.08}$ & 1.965$_{-0.190}^{+0.136}$ & 2.24$_{-2.19}^{+1.85}$ & 44.33$_{-0.08}^{+0.08}$ & 3.33$_{-0.32}^{+0.69}$ & 0.904 & 44.65$_{-0.23}^{+0.03}$ & 2.111$_{-0.351}^{+0.089}$ & 1.99$_{-1.99}^{+1.05}$ & 44.12$_{-0.13}^{+0.29}$ & >0.48$_{}^{}$ & 2.90$_{-0.71}^{+0.39}$ & -0.22$_{-0.19}^{+0.65}$ & 0.905 \\ 
36 & 44.45$_{-0.01}^{+0.01}$ & 2.264$_{-0.034}^{+0.038}$ & >4.35$_{}^{}$ & 43.02$_{-0.14}^{+0.30}$ & 0.08$_{-0.02}^{+0.03}$ & 1.103 & 44.45$_{-0.02}^{+0.01}$ & 2.243$_{-0.063}^{+0.040}$ & 4.85$_{-3.73}^{+4.85}$ & 43.15$_{-0.16}^{+0.23}$ & 5.90$_{-1.86}^{+3.13}$ & 1.101 & 44.45$_{-0.02}^{+0.01}$ & 2.249$_{-0.060}^{+0.067}$ & >3.95$_{}^{}$ & 43.13$_{-0.24}^{+0.23}$ & >0.16$_{}^{}$ & 6.10$_{-3.61}^{+3.18}$ & -1.03$_{-0.25}^{+0.25}$ & 1.103 \\ 
37 & 45.05$_{-0.01}^{+0.01}$ & 2.403$_{-0.051}^{+0.035}$ & >4.83$_{}^{}$ & 43.80$_{-0.12}^{+0.11}$ & 0.08$_{-0.03}^{+0.03}$ & 1.069 & 45.03$_{-0.06}^{+0.02}$ & 2.334$_{-0.123}^{+0.068}$ & >3.06$_{}^{}$ & 44.42$_{-0.34}^{+0.18}$ & 4.71$_{-1.25}^{+4.71}$ & 1.059 & 45.03$_{-0.06}^{+0.02}$ & 2.340$_{-0.118}^{+0.133}$ & 5.00$_{-1.29}^{+5.0}$ & 44.04$_{-0.21}^{+0.23}$ & >0.16$_{}^{}$ & 4.81$_{-1.51}^{+2.56}$ & -0.73$_{-0.24}^{+0.42}$ & 1.062 \\ 
38 & 44.27$_{-0.01}^{+0.02}$ & 2.140$_{-0.072}^{+0.080}$ & 4.37$_{-3.69}^{+4.37}$ & 43.12$_{-0.33}^{+0.41}$ & 0.13$_{-0.02}^{+0.03}$ & 0.887 & 44.24$_{-0.07}^{+0.04}$ & 1.993$_{-0.210}^{+0.139}$ & 2.28$_{-2.28}^{+2.28}$ & 44.04$_{-0.66}^{+0.38}$ & 4.81$_{-1.90}^{+1.04}$ & 0.889 & 44.27$_{-0.02}^{+0.03}$ & 2.110$_{-0.096}^{+0.105}$ & 3.59$_{-3.6}^{+3.6}$ & 43.34$_{-0.50}^{+0.96}$ & 0.20$_{-0.1}^{+0.64}$ & <4.61$_{}^{}$ & -0.60$_{-0.42}^{+0.95}$ & 0.889 \\ 
39 & 44.33$_{-0.03}^{+0.03}$ & 2.571$_{-0.126}^{+0.141}$ & 2.03$_{-2.03}^{+2.03}$ & 43.15$_{-0.51}^{+0.25}$ & 0.11$_{-0.05}^{+0.02}$ & 0.865 & 44.11$_{-0.50}^{+0.11}$ & 2.176$_{-0.000}^{+0.000}$ & 0.00$_{-0.0}^{+4.55}$ & 43.93$_{-0.85}^{+0.22}$ & 3.30$_{-0.00}^{+0.00}$ & 0.858 & 44.28$_{-0.42}^{+0.04}$ & 2.399$_{-0.645}^{+0.136}$ & 0.00$_{-0.0}^{+4.1}$ & 43.57$_{-0.33}^{+0.55}$ & >0.16$_{}^{}$ & 2.79$_{-2.49}^{+1.73}$ & -0.50$_{-0.40}^{+1.25}$ & 0.857 \\ 
40 & 44.79$_{-0.00}^{+0.00}$ & 2.364$_{-0.020}^{+0.022}$ & >4.89$_{}^{}$ & 43.59$_{-0.07}^{+0.06}$ & 0.12$_{-0.01}^{+0.01}$ & 1.047 & 44.51$_{-0.09}^{+0.13}$ & 1.810$_{-0.079}^{+0.185}$ & 2.23$_{-0.43}^{+0.23}$ & 44.42$_{-0.10}^{+0.05}$ & 2.95$_{-0.11}^{+0.41}$ & 1.028 & 44.53$_{-0.10}^{+0.19}$ & 1.837$_{-0.131}^{+0.138}$ & 2.28$_{-0.9}^{+0.66}$ & 44.41$_{-0.30}^{+0.06}$ & >1.33$_{}^{}$ & 2.97$_{-0.13}^{+0.17}$ & 0.30$_{-0.57}^{+0.23}$ & 1.029 \\ 
41 & 43.62$_{-0.01}^{+0.01}$ & 1.754$_{-0.042}^{+0.042}$ & 0.03$_{-0.03}^{+2.63}$ & 42.37$_{-0.13}^{+0.10}$ & 0.14$_{-0.02}^{+0.02}$ & 1.050 & 43.54$_{-0.19}^{+0.06}$ & 1.598$_{-0.279}^{+0.133}$ & <1.79$_{}^{}$ & 42.83$_{-0.15}^{+0.27}$ & 2.82$_{-0.48}^{+1.06}$ & 1.050 & 43.61$_{-0.32}^{+0.02}$ & 1.684$_{-0.578}^{+0.080}$ & 0.01$_{-0.01}^{+1.93}$ & 42.57$_{-0.10}^{+0.57}$ & >0.49$_{}^{}$ & 1.78$_{-3.85}^{+1.88}$ & -0.53$_{-0.29}^{+0.53}$ & 1.030 \\ 
42 & 44.64$_{-0.01}^{+0.01}$ & 2.019$_{-0.030}^{+0.034}$ & 4.72$_{-0.86}^{+4.72}$ & 43.28$_{-0.13}^{+0.20}$ & 0.11$_{-0.03}^{+0.02}$ & 1.063 & 44.48$_{-0.05}^{+0.11}$ & 1.520$_{-0.102}^{+0.222}$ & 1.01$_{-1.02}^{+1.95}$ & 43.92$_{-0.24}^{+0.18}$ & 2.78$_{-0.10}^{+0.60}$ & 1.050 & 44.56$_{-0.19}^{+0.04}$ & 1.593$_{-0.319}^{+0.114}$ & 0.17$_{-0.17}^{+1.27}$ & 43.95$_{-0.15}^{+0.16}$ & >0.65$_{}^{}$ & 2.69$_{-0.72}^{+0.48}$ & -0.11$_{-0.23}^{+0.49}$ & 1.048 \\ 
43 & 44.03$_{-0.01}^{+0.01}$ & 2.215$_{-0.068}^{+0.065}$ & 3.79$_{-0.59}^{+0.47}$ & 43.39$_{-0.15}^{+0.16}$ & 0.09$_{-0.00}^{+0.01}$ & 1.034 & 43.99$_{-0.02}^{+0.02}$ & 2.068$_{-0.078}^{+0.071}$ & <0.04$_{}^{}$ & 43.60$_{-0.21}^{+0.25}$ & 4.98$_{-0.55}^{+0.53}$ & 1.089 & 44.03$_{-0.01}^{+0.01}$ & 2.254$_{-0.065}^{+0.067}$ & 2.35$_{-0.59}^{+0.62}$ & 43.25$_{-0.03}^{+0.03}$ & 0.08$_{-0.0}^{+0.02}$ & <-1.64$_{}^{}$ & -0.39$_{-0.14}^{+0.15}$ & 1.033 \\ 
44 & 44.92$_{-0.02}^{+0.03}$ & 2.086$_{-0.071}^{+0.089}$ & >3.45$_{}^{}$ & 44.05$_{-0.45}^{+0.54}$ & 0.13$_{-0.02}^{+0.04}$ & 0.850 & 44.85$_{-0.04}^{+0.02}$ & 1.726$_{-0.476}^{+0.086}$ & 0.01$_{-0.01}^{+1.07}$ & 44.73$_{-0.38}^{+0.27}$ & 4.55$_{-1.59}^{+1.42}$ & 0.845 & 44.86$_{-0.21}^{+0.03}$ & 1.729$_{-0.251}^{+0.105}$ & 0.03$_{-0.03}^{+1.02}$ & 44.18$_{-0.19}^{+0.31}$ & 0.51$_{-0.37}^{+160.61}$ & 1.70$_{-4.0}^{+1.77}$ & -0.23$_{-0.25}^{+0.56}$ & 0.844 \\ 
45 & 44.67$_{-0.01}^{+0.01}$ & 2.411$_{-0.035}^{+0.044}$ & >4.76$_{}^{}$ & 43.22$_{-0.14}^{+0.33}$ & 0.07$_{-0.03}^{+0.02}$ & 1.050 & 44.61$_{-0.10}^{+0.03}$ & 2.207$_{-0.179}^{+0.100}$ & 0.77$_{-0.77}^{+0.9}$ & 43.78$_{-0.27}^{+0.49}$ & 4.08$_{-0.85}^{+1.42}$ & 1.042 & 44.65$_{-0.04}^{+0.01}$ & 2.333$_{-0.102}^{+0.056}$ & 2.95$_{-1.87}^{+1.78}$ & 43.49$_{-0.53}^{+0.32}$ & >0.16$_{}^{}$ & 5.13$_{-2.82}^{+2.39}$ & -0.91$_{-0.25}^{+0.39}$ & 1.046 \\ 
46 & 44.50$_{-0.00}^{+0.01}$ & 2.586$_{-0.024}^{+0.022}$ & 4.74$_{-1.58}^{+4.74}$ & 42.92$_{-0.16}^{+0.13}$ & 0.07$_{-0.03}^{+0.03}$ & 1.047 & 44.36$_{-0.23}^{+0.09}$ & 2.263$_{-0.269}^{+0.144}$ & <0.97$_{}^{}$ & 44.14$_{-0.32}^{+0.10}$ & 3.40$_{-0.43}^{+0.82}$ & 1.039 & 44.37$_{-0.24}^{+0.60}$ & 2.270$_{-0.272}^{+0.155}$ & 0.01$_{-0.01}^{+0.79}$ & 43.88$_{-0.42}^{+0.28}$ & >0.47$_{}^{}$ & 3.42$_{-0.97}^{+0.81}$ & -0.24$_{-0.55}^{+0.60}$ & 1.040 \\ 
47 & 43.88$_{-0.01}^{+0.01}$ & 2.401$_{-0.060}^{+0.061}$ & >2.37$_{}^{}$ & 42.59$_{-0.19}^{+0.31}$ & 0.09$_{-0.02}^{+0.02}$ & 1.119 & 43.85$_{-0.09}^{+0.03}$ & 2.276$_{-0.195}^{+0.108}$ & 3.33$_{-3.33}^{+3.33}$ & 42.97$_{-0.31}^{+0.34}$ & 4.25$_{-0.95}^{+1.64}$ & 1.114 & 43.83$_{-0.16}^{+0.05}$ & 2.193$_{-0.246}^{+0.123}$ & 1.35$_{-1.34}^{+1.34}$ & 43.04$_{-0.27}^{+0.23}$ & >0.17$_{}^{}$ & 3.35$_{-2.55}^{+1.55}$ & -0.51$_{-0.33}^{+0.29}$ & 1.116 \\ 
48 & 44.51$_{-0.01}^{+0.01}$ & 2.293$_{-0.044}^{+0.044}$ & 5.00$_{-2.1}^{+5.0}$ & 43.18$_{-0.20}^{+0.16}$ & 0.12$_{-0.04}^{+0.03}$ & 1.026 & 44.21$_{-0.27}^{+0.18}$ & 1.592$_{-0.442}^{+0.298}$ & 0.01$_{-0.01}^{+3.18}$ & 44.14$_{-0.14}^{+0.08}$ & 2.89$_{-1.68}^{+1.10}$ & 1.011 & 44.38$_{-0.37}^{+0.06}$ & 1.822$_{-0.633}^{+0.151}$ & 0.02$_{-0.02}^{+2.16}$ & 43.94$_{-0.18}^{+0.23}$ & >0.52$_{}^{}$ & 2.60$_{-0.89}^{+0.47}$ & -0.05$_{-0.29}^{+0.64}$ & 1.013 \\ 
49 & 44.08$_{-0.01}^{+0.01}$ & 2.730$_{-0.072}^{+0.071}$ & >4.66$_{}^{}$ & 43.08$_{-0.19}^{+0.17}$ & 0.11$_{-0.02}^{+0.01}$ & 0.967 & 43.60$_{-0.19}^{+0.32}$ & 1.723$_{-0.465}^{+0.461}$ & 1.23$_{-1.45}^{+3.39}$ & 43.11$_{-0.45}^{+0.50}$ & 3.23$_{-0.17}^{+0.29}$ & 0.942 & 43.60$_{-0.19}^{+0.40}$ & 1.760$_{-0.466}^{+0.540}$ & 1.56$_{-1.56}^{+2.23}$ & 43.91$_{-0.42}^{+0.05}$ & >0.46$_{}^{}$ & 3.23$_{-0.48}^{+1.03}$ & 0.73$_{-1.06}^{+0.56}$ & 0.939 \\ 
50 & 45.12$_{-0.02}^{+0.02}$ & 1.944$_{-0.073}^{+0.085}$ & 1.95$_{-1.97}^{+1.97}$ & 44.20$_{-0.33}^{+0.46}$ & 0.14$_{-0.02}^{+0.03}$ & 0.917 & 45.11$_{-0.14}^{+0.04}$ & 1.884$_{-0.416}^{+0.125}$ & 1.18$_{-1.06}^{+1.06}$ & 45.22$_{-0.98}^{+0.50}$ & 5.40$_{-2.86}^{+1.18}$ & 0.923 & 45.12$_{-0.03}^{+0.04}$ & 1.939$_{-0.056}^{+0.099}$ & 1.87$_{-1.87}^{+1.87}$ & 44.00$_{-0.15}^{+0.23}$ & 0.12$_{-0.02}^{+0.29}$ & <2.61$_{}^{}$ & -0.63$_{-0.28}^{+1.43}$ & 0.920 \\ 
51 & 44.05$_{-0.00}^{+0.01}$ & 2.014$_{-0.015}^{+0.030}$ & 4.16$_{-1.16}^{+4.16}$ & 42.23$_{-0.15}^{+0.38}$ & 0.08$_{-0.02}^{+0.02}$ & 1.077 & 44.03$_{-0.03}^{+0.01}$ & 1.906$_{-0.064}^{+0.038}$ & 2.30$_{-2.31}^{+2.3}$ & 42.89$_{-0.34}^{+0.36}$ & 3.84$_{-0.75}^{+1.06}$ & 1.073 & 44.02$_{-0.05}^{+0.02}$ & 1.859$_{-0.091}^{+0.056}$ & 1.71$_{-1.71}^{+1.01}$ & 42.99$_{-0.35}^{+0.31}$ & >0.39$_{}^{}$ & 3.61$_{-1.36}^{+0.71}$ & -0.62$_{-0.39}^{+0.33}$ & 1.074 \\ 
52 & 45.09$_{-0.00}^{+0.01}$ & 2.365$_{-0.008}^{+0.008}$ & 2.95$_{-0.09}^{+0.09}$ & 44.11$_{-0.02}^{+0.02}$ & 0.14$_{-0.00}^{+0.00}$ & 1.089 & 44.95$_{-0.03}^{+0.02}$ & 1.976$_{-0.045}^{+0.038}$ & <0.04$_{}^{}$ & 44.87$_{-0.03}^{+0.03}$ & 3.93$_{-0.10}^{+0.20}$ & 1.084 & 45.05$_{-0.01}^{+0.01}$ & 2.211$_{-0.016}^{+0.018}$ & 0.99$_{-0.23}^{+0.27}$ & 44.37$_{-0.03}^{+0.06}$ & 0.36$_{-0.06}^{+0.06}$ & 1.31$_{-0.4}^{+0.45}$ & -0.41$_{-0.04}^{+0.04}$ & 1.064 \\ 
53 & 44.98$_{-0.01}^{+0.01}$ & 1.832$_{-0.070}^{+0.024}$ & 1.56$_{-0.77}^{+0.74}$ & 42.81$_{-2.51}^{+0.47}$ & 0.06$_{-0.05}^{+0.23}$ & 1.070 & 44.86$_{-0.36}^{+0.26}$ & 1.547$_{-0.578}^{+0.139}$ & <2.24$_{}^{}$ & 44.37$_{-1.65}^{+0.14}$ & 2.53$_{-2.53}^{+1.55}$ & 1.062 & 44.93$_{-0.83}^{+0.03}$ & 1.623$_{-1.202}^{+0.083}$ & <0.80$_{}^{}$ & 43.94$_{-0.32}^{+0.51}$ & >0.47$_{}^{}$ & 1.99$_{-1.21}^{+0.54}$ & -0.52$_{-0.38}^{+1.87}$ & 1.063 \\ 
54 & 45.88$_{-0.01}^{+0.01}$ & 2.018$_{-0.032}^{+0.041}$ & 3.04$_{-0.56}^{+1.4}$ & 44.62$_{-0.15}^{+0.10}$ & 0.16$_{-0.02}^{+0.02}$ & 0.922 & 45.82$_{-0.20}^{+0.04}$ & 1.815$_{-0.267}^{+0.113}$ & 1.42$_{-1.01}^{+0.66}$ & 45.49$_{-0.24}^{+0.35}$ & 3.93$_{-1.34}^{+1.03}$ & 0.918 & 45.84$_{-0.20}^{+0.03}$ & 1.865$_{-0.274}^{+0.077}$ & 1.56$_{-0.76}^{+0.84}$ & 45.31$_{-0.68}^{+0.54}$ & >0.30$_{}^{}$ & 3.20$_{-3.95}^{+1.78}$ & -0.13$_{-0.63}^{+0.50}$ & 0.919 \\ 
55 & 43.35$_{-0.00}^{+0.00}$ & 1.684$_{-0.015}^{+0.016}$ & 2.96$_{-0.95}^{+0.97}$ & 41.63$_{-0.09}^{+0.20}$ & 0.10$_{-0.01}^{+0.01}$ & 1.070 & 43.29$_{-0.06}^{+0.03}$ & 1.406$_{-0.112}^{+0.073}$ & 0.38$_{-0.38}^{+0.7}$ & 42.41$_{-0.16}^{+0.16}$ & 2.70$_{-0.31}^{+0.38}$ & 1.060 & 43.29$_{-0.06}^{+0.04}$ & 1.405$_{-0.115}^{+0.099}$ & 0.35$_{-0.35}^{+1.15}$ & 42.41$_{-0.26}^{+0.17}$ & >0.91$_{}^{}$ & 2.69$_{-0.3}^{+0.37}$ & -0.29$_{-0.35}^{+0.29}$ & 1.061 \\ 
56 & 43.09$_{-0.01}^{+0.01}$ & 2.894$_{-0.046}^{+0.047}$ & >4.94$_{}^{}$ & 42.90$_{-0.02}^{+0.02}$ & 0.11$_{-0.00}^{+0.00}$ & 1.301 & 42.74$_{-0.03}^{+0.02}$ & 1.665$_{-0.096}^{+0.091}$ & <0.07$_{}^{}$ & 43.37$_{-0.04}^{+0.04}$ & 4.77$_{-0.15}^{+0.15}$ & 1.348 & 42.99$_{-0.02}^{+0.02}$ & 2.478$_{-0.088}^{+0.096}$ & 2.95$_{-1.0}^{+1.65}$ & 43.17$_{-0.07}^{+0.06}$ & 0.28$_{-0.04}^{+0.05}$ & 2.21$_{-0.48}^{+0.42}$ & 0.39$_{-0.08}^{+0.08}$ & 1.119 \\ 
57 & 43.74$_{-0.01}^{+0.02}$ & 1.934$_{-0.052}^{+0.101}$ & 2.78$_{-1.1}^{+1.66}$ & 41.09$_{-41.09}^{+1.62}$ & 0.03$_{-0.03}^{+0.03}$ & 0.949 & 43.74$_{-1.54}^{+0.02}$ & 1.936$_{-0.168}^{+0.079}$ & 3.10$_{-1.36}^{+1.41}$ & 42.63$_{-42.63}^{+0.61}$ & 4.16$_{-6.11}^{+4.16}$ & 0.948 & 43.70$_{-0.72}^{+0.04}$ & 1.785$_{-1.785}^{+1.785}$ & 2.30$_{-1.1}^{+2.54}$ & 42.95$_{-42.95}^{+0.36}$ & >0.16$_{}^{}$ & <11.00$_{}^{}$ & -0.33$_{-4.56}^{+0.45}$ & 0.950 \\ 
58 & 44.73$_{-0.01}^{+0.01}$ & 1.981$_{-0.037}^{+0.041}$ & >0.00$_{}^{}$ & 43.65$_{-0.09}^{+0.07}$ & 0.15$_{-0.01}^{+0.01}$ & 1.099 & 44.57$_{-0.27}^{+0.08}$ & 1.630$_{-0.266}^{+0.057}$ & >0.00$_{}^{}$ & 44.29$_{-0.05}^{+0.06}$ & 2.96$_{-0.57}^{+0.26}$ & 1.072 & 44.68$_{-0.10}^{+0.02}$ & 1.799$_{-0.053}^{+0.053}$ & >0.00$_{}^{}$ & 44.02$_{-0.13}^{+0.13}$ & 1.06$_{-0.55}^{+2.14}$ & 2.26$_{-0.34}^{+0.45}$ & -0.24$_{-0.19}^{+0.39}$ & 1.070 \\ 
59 & 43.07$_{-0.01}^{+0.01}$ & 1.963$_{-0.024}^{+0.025}$ & 3.88$_{-1.4}^{+0.31}$ & 41.47$_{-0.16}^{+0.24}$ & 0.10$_{-0.02}^{+0.02}$ & 0.973 & 43.03$_{-0.08}^{+0.02}$ & 1.797$_{-0.146}^{+0.072}$ & 1.84$_{-1.84}^{+2.7}$ & 42.06$_{-0.25}^{+0.29}$ & 3.31$_{-0.60}^{+0.79}$ & 0.968 & 43.05$_{-0.09}^{+0.01}$ & 1.849$_{-0.164}^{+0.052}$ & 2.05$_{-1.24}^{+0.34}$ & 41.88$_{-0.22}^{+0.44}$ & >0.18$_{}^{}$ & 3.05$_{-2.73}^{+1.17}$ & -0.76$_{-0.25}^{+0.59}$ & 0.969 \\ 
\enddata
\tablecomments{
The spectral fitting results for model 1, 2 and 3 (see \S \ref{model}), including: 1) Primary continuum luminosity ($L_\text{PC,bb},L_\text{PC,po},L_\text{PC,cpl}$) in logarithmic scale, photon index ($\Gamma_\text{PC,bb},\Gamma_\text{PC,po},\Gamma_\text{PC,cpl}$), and reflection strength ($R_\text{bb},R_\text{po},R_\text{cpl}$). 2) Soft excess luminosity ($L_\text{SE,0.5-2,bb},L_\text{SE,0.5-2,po},L_\text{SE,0.5-2,cpl}$) in logarithmic scale, ``shape'' parameter ($T_\text{bb}$ for model 1,$\Gamma_\text{SE,po}$ for model 2, and $E_\text{cut}, \Gamma_\text{SE,cpl}$ for model 3), and soft excess strength ($\log q$, see Eq. \ref{q}) under model 3. 3) Reduced chi-square ($\chi^2_{\nu,\text{bb}},\chi^2_{\nu,\text{po}},\chi^2_{\nu,\text{cpl}}$). See text for more details.
}
\end{deluxetable*}
\end{longrotatetable}


\bibliography{XMMSE_profile}{}
\bibliographystyle{aasjournal}

\end{document}